\documentclass[11pt]{article}

\usepackage[utf8]{inputenc}
\usepackage[T1]{fontenc}
\usepackage{lmodern}
\usepackage[a4paper,margin=2.5cm]{geometry}
\usepackage{amsmath,amssymb}
\usepackage{graphicx}
\usepackage{subfig}
\usepackage{bm}
\usepackage{cancel}
\usepackage[round,authoryear]{natbib}
\usepackage{accents}
\usepackage{tikz,xcolor}
\usepackage{float}
\usepackage[colorlinks=true,linkcolor=blue,citecolor=blue,urlcolor=blue]{hyperref}

\title{Transient coupling of Marangoni spreading and interfacial waves}
\author{Abhinandan Mohanty and Palas Kumar Farsoiya\thanks{Corresponding author: palas.farsoiya@ch.iitr.ac.in}\\[4pt]
\small Department of Chemical Engineering, Indian Institute of Technology Roorkee,\\
\small Roorkee, Uttarakhand 247667, India}
\date{}
\newcommand{\UP}{_u}
\newcommand{\LO}{_l}
\newcommand{\Rey}{\mathit{Re}}
\newcommand{\Oh}{\mathit{Oh}}
\newcommand{\Bo}{\mathit{Bo}}
\newcommand{\Pes}{\mathit{Pe}_s}
\newcommand{\tr}{\tilde t}

\newcommand{\yr}{\tilde y}
\newcommand{\NN}{\mathcal{N}}
\newcommand{\Nc}{\mathcal{N}_{\!c}}
\newcommand{\QQ}{\mathcal{Q}}
\newcommand{\Aa}{\mathcal{A}}
\newcommand{\B}{\mathcal{B}}
\newcommand{\Cc}{\mathcal{C}}

\newcommand{\erfc}{\mathrm{erfc}\,}

\newcommand{\Lap}{\mathcal{L}}
\newcommand{\Gb}{\breve\Gamma}
\newcommand{\Gh}{\hat\Gamma}
\newcommand{\eb}{\breve\eta}
\newcommand{\eh}{\hat\eta}
\def\u{\mathbf{u}}

\begin{document}
\maketitle

\begin{abstract}
How does an initially non-uniform distribution of surfactant generate waves on a deformable interface, and how do those waves then govern the redistribution of the surfactant? We answer this question for the linear Cauchy-Poisson problem of an axisymmetric interface between two viscous fluids, released from rest with independently prescribed distributions of surfactant and of interface displacement, by Hankel and Laplace transforms. In Stokes flow the pressure and the viscous normal stress of the Marangoni flow cancel in the normal traction on the interface, for any viscosity ratio, so that the surfactant relaxes without displacing the interface: the coupling between surfactant redistribution and interface deformation is inertial. With inertia, the Marangoni flow initiates a capillary-gravity wave whose polarity, at leading order for thin boundary layers, is set by the sign of $\rho_l\mu_l-\rho_u\mu_u$, and the surfactant is then carried by the wave, so that its concentration performs weakly damped oscillations at the wave frequency after the Marangoni relaxation of a flat surface has ended. The amplitude of these oscillations is given in closed form at small viscosity; its maximum does not coincide with that of the familiar resonant damping, and the surfactant and the interface shape decouple exactly only when the two fluids have equal densities and viscosities. A localized displacement radiates a wave packet that the surfactant damps, and a localized surfactant redistribution initiates a wave on an initially flat interface. Direct numerical simulations with sharp geometric surfactant transport verify these predictions.
\end{abstract}


\section{Introduction}
\label{sec:intro}

Surface-active contaminants can alter interfacial waves far more strongly than their effect on the mean surface tension would suggest. Gradients of surfactant concentration generated by wave motion produce Marangoni stresses that resist surface dilation and modify viscous dissipation near the interface \citep{levich1962physicochemical,lucassen1968longitudinal}. A surfactant-covered interface supports both transverse capillary-gravity waves and longitudinal Marangoni waves, whose restoring force is supplied by surface elasticity. Their interaction can produce maximum wave damping at a finite elasticity \citep{lucassen1970properties,Rajan2020resonance}.

For clean interfaces, \citet{prosperetti1976,prosperetti1981motion} formulated the small-amplitude motion of one and two viscous fluids as initial-value problems (IVPs). The analysis was extended to multiple interfaces by \citet{VartdalMultiIVP2019}, who obtained coupled equations for the Laplace-transformed interface amplitudes and examined their time-dependent response. Related axisymmetric, azimuthal and finite-depth problems have also been investigated \citep{farsoiya2017axisymmetric,farsoiya2019free,farsoiya2020azimuthal,dasgupta2023}. These studies establish that the evolution from a prescribed initial state involves the diffusion of vorticity as well as discrete wave modes. For an infinitely deep fluid, this diffusive contribution is represented by a branch cut in the Laplace plane. With an insoluble surfactant present, \citet{Denner2017onefluid} derived and solved a coupled integro-differential IVP for a single viscous fluid, including surface advection and diffusion, and proposed a correction to the critical damping wavelength in a diffusion-dominated regime. Their analytical inversion involves complementary error functions and retains viscous memory; it is not merely a finite sum of exponentially evolving normal modes.

The modal dynamics of a surfactant-covered interface between two viscous fluids are also well established. \citet{RajanHenderson2018} derived a comprehensive dispersion relation for fluids of arbitrary densities and viscosities and analysed the dissipation of capillary-gravity waves. \citet{Rajan2020resonance} investigated the coupled Marangoni branch and showed that resonance can maximise transverse-wave damping while minimising longitudinal-wave damping. \citet{Rajan2021} further examined the two branches, the transition between overdamped and underdamped motion, and extrema of frequency and damping rate. Resonance damping has also been revisited for films of finite thickness \citep{ErmakovKhazanov2022}. The interaction of the longitudinal and transverse branches, and the resonant damping of capillary-gravity waves that it produces, are therefore established; they are properties of the normal modes, obtained from the dispersion relation, and they are not the subject of the present work. What a dispersion relation cannot give is the transient: how an initial disturbance distributes itself among the modes and the continuous spectrum, and how surfactant redistribution and interface deformation drive each other in time.

Unsteady Marangoni spreading from a localized surfactant distribution was examined by \citet{Jensen1995spreading}, who coupled insoluble-surfactant transport to a viscous boundary-layer flow and allowed the free surface to deform. For a fixed surfactant mass on an initially clean, deep liquid, the similarity solutions predict spreading lengths proportional to $t^{1/2}$ for a planar strip and $t^{3/8}$ for an axisymmetric patch. The study also established the influence of background contamination and proposed a reduced model for the transition to finite-depth spreading; verification against full unsteady Navier-Stokes solutions was left for future work.

Marangoni spreading on a deep layer has since been analysed in several other limits, all with the interface held flat. For a surfactant supplied steadily at a point, \citet{Mandre2017spreading} obtained boundary-layer similarity solutions in which the surface velocity decays with distance as $r^{-3/5}$ when the surfactant spreads in the adsorbed phase and as $r^{-1}$ when it spreads in the dissolved phase. In the opposite, Stokes limit the transport problem is nonlocal but can be mapped onto a complex Burgers equation, from which \citet{Bickel2022exact} obtained explicit solutions for pulses, holes and periodic distributions, including a circular pulse whose radius grows as $t^{1/3}$, and \citet{TempranoColeto2024self} identified six self-similar solutions, with exponent $1/2$ for spreading pulses and $2$ or $3/2$ for depleted regions that fill. These studies are nonlinear, but they also delimit the linear regime considered here. When the concentration perturbation is small compared with a pre-existing uniform concentration, the transport equations become linear, and a contaminated interface is then not equivalent to a clean interface of reduced tension \citep{Bickel2022exact}. In Stokes flow a sinusoidal perturbation decays exponentially \citep{Bickel2022exact,McNair2022cavity}. \citet{Shardt2016oscillatory} solved the corresponding linear initial-value problem with inertia, for a single deep liquid released from rest, by Laplace transform. Above a critical Reynolds number the poles of their solution become complex and the concentration overshoots uniformity, but it changes sign at most three times before decaying monotonically as $t^{-3/2}$, a behaviour that distinguishes it from a damped harmonic oscillator; the decay time scales as $(\rho\lambda^4\mu/\beta^2)^{1/3}$ for a perturbation of wavelength $\lambda$, and the flow beneath the surface alternates between one and two rows of counter-rotating vortices. In all of these analyses, however, the interface is held flat and bounds a single fluid. \citet{Shardt2016oscillatory} identified a deforming surface and two viscous fluids as natural extensions, and the capillary-gravity waves that Marangoni stresses initiate on a deformable interface, observed during spreading on a deep subphase \citep{Sauleda2022spreading}, lie outside the scope of these studies.

The present paper takes up these two extensions. We consider an axisymmetric interface between two superposed viscous fluids of arbitrary densities and viscosities, with surface tension and gravity, released from rest with an arbitrary small-amplitude distribution of surfactant and an independently prescribed displacement. This is the Cauchy-Poisson problem of water-wave theory, in which the free surface is released from a prescribed shape \citep{lamb1932hydrodynamics,prosperetti1976}, extended to a surfactant-laden interface, for which the prescribed initial state includes the distribution of surfactant. The linear problem is solved by Hankel and Laplace transforms in the scaling of \citet{Shardt2016oscillatory}, in which the Marangoni stress sets the velocity scale and a single Reynolds number $\Rey=\rho\LO\beta\lambda/\mu\LO^2$ measures inertia, so that the two solutions can be compared directly; the deformable interface brings in the elasticity number $E=\beta/\sigma_0$, a Bond number and the density and viscosity ratios.

The central question is how an initially non-uniform surfactant distribution generates interfacial waves, and how those waves subsequently govern the redistribution of the surfactant. The answer has five parts. (i) In Stokes flow the pressure and the viscous normal stress of the Marangoni flow cancel exactly on the interface, for any viscosity ratio, so that the surfactant relaxes without displacing the interface; the coupling is an inertial effect. (ii) With inertia, the difference between the shear stresses of the two vortical boundary layers, measured at leading order for thin layers by $\chi=(1-\phi)/(1+\phi)$ with $\phi=(\rho\UP\mu\UP/\rho\LO\mu\LO)^{1/2}$, initiates a capillary-gravity wave, whose polarity reverses with the sign of $\chi$. (iii) The surfactant is then carried by the wave, and its concentration performs weakly damped oscillations at the wave frequency after the Marangoni relaxation of a flat surface would have ended; we separate the contributions of the Marangoni poles, of the wave poles and of the continuous spectrum, and identify three temporal regimes. (iv) The amplitude of these oscillations, the residue of the solution at the wave pole, is given at small viscosity by a closed-form expression, whose maximum, whether the Reynolds number or the elasticity is varied, differs from the maximum of the damping; surfactant and shape decouple exactly only when the two fluids have equal densities and viscosities. (v) For localized disturbances, retaining the whole spectrum by the Hankel integral, a localized displacement radiates a wave packet that the surfactant damps, and a localized surfactant redistribution initiates a wave on an initially flat interface.

The paper is organised as follows. Section~\ref{sec:problem} states the problem and its non-dimensional form, and \S\ref{sec:analytical} gives the Hankel-Laplace solution and its inversion. Section~\ref{sec:single} analyses a single mode: the poles, the Stokes limit, the limit of large Reynolds number, and the temporal regimes. Section~\ref{sec:num} summarises the direct numerical simulations, whose verification is reported in the supplementary material, and \S\ref{sec:dns-results} compares them with the single-mode solution. Section~\ref{sec:deposit} treats localized disturbances, and \S\ref{sec:conclusions} concludes.

\section{Problem definition}
\label{sec:problem}

\subsection{Governing equations}

\begin{figure}
	\centering
	\includegraphics[width=\textwidth]{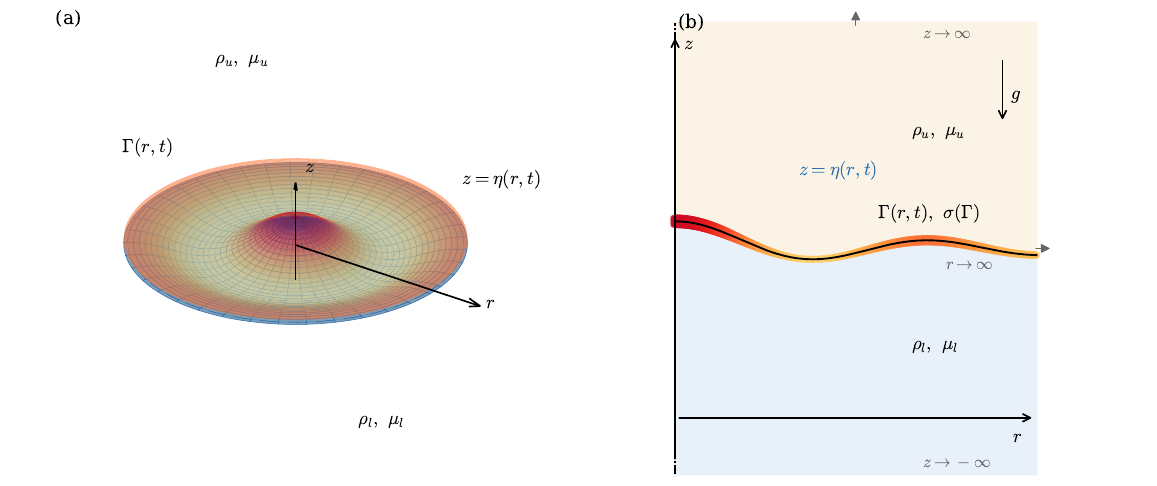}
	\caption{Configuration. (a) The axisymmetric interface $z=\eta(r,t)$ between the lower fluid ($\rho\LO$, $\mu\LO$) and the upper fluid ($\rho\UP$, $\mu\UP$), carrying an insoluble surfactant of concentration $\Gamma(r,t)$ (coloured layer, whose shade and thickness are proportional to $\Gamma$), for the Bessel mode $J_0(kr)$. (b) The axisymmetric $(r,z)$ section: both fluids are unbounded in $r$ and in $z$, gravity acts along $-z$, and the surface tension $\sigma(\Gamma)$ follows \eqref{eos}. Both fluids are at rest initially, and the initial distributions $\eta(r,0)$ and $\Gamma(r,0)$ are prescribed independently.}
	\label{fig:config}
\end{figure}

Two viscous, incompressible and immiscible fluids are separated by an interface which is the plane $z=0$ in the base state (figure~\ref{fig:config}). We use cylindrical coordinates $(r,\vartheta,z)$ and consider axisymmetric motion without swirl, so that the velocity is $\mathbf u=(u_r,u_z)$ and nothing depends on $\vartheta$. The lower fluid ($z<0$) has density $\rho\LO$ and dynamic viscosity $\mu\LO$; the upper fluid ($z>0$) has $\rho\UP$ and $\mu\UP$. Gravity $g$ acts along $-z$ and the configuration is stable for $\rho\LO>\rho\UP$. Both fluids are unbounded in $r$ and in $z$, and disturbances vanish as $r\to\infty$ and as $z\to\pm\infty$. The lower fluid is the liquid of \citet{Shardt2016oscillatory}, whose problem is planar; their $y$ axis points into the liquid, our $z$ axis points out of it.

The interface, $z=\eta(r,t)$, carries an insoluble surfactant of concentration $\Gamma(r,t)$ and surface diffusivity $D_s$, which obeys \citep{stone1990simple}
\begin{equation}
  \frac{\partial \Gamma}{\partial t} + \nabla_s\cdot(\Gamma\,\mathbf{u}_s) + \Gamma\,(\nabla_s\cdot\mathbf{n})(\mathbf{u}\cdot\mathbf{n}) = D_s\nabla_s^2\Gamma,
  \label{eq:surf-full}
\end{equation}
where $\mathbf{u}_s=(\mathsf I-\mathbf n\mathbf n)\cdot\mathbf u$ is the tangential component of the interfacial velocity, $\mathbf n$ the unit normal, pointing into the upper fluid, and $\nabla_s$ the surface gradient; on the flat base interface $\nabla_s\cdot\mathbf u_s=r^{-1}\partial(ru_r)/\partial r$. The third term, absent from the flat-surface problem, is the change of concentration produced by the dilatation of a curved interface that moves normally to itself. The base state has a uniform concentration $\Gamma_0$ and surface tension $\sigma_0$, and the surface tension follows the linear equation of state \citep{timmermans2002effect}
\begin{equation}
  \sigma=\sigma_0-\beta\left(\frac{\Gamma}{\Gamma_0}-1\right),
  \label{eos}
\end{equation}
where $\beta=-\Gamma_0\,(\partial\sigma/\partial\Gamma)_{\Gamma_0}$ is the Gibbs elasticity. Equation \eqref{eos} is the equation of state $\gamma=\gamma_0-\alpha\Gamma$ of \citet{Shardt2016oscillatory} written about the base state, with $\beta=\alpha\Gamma_0$ and $\sigma_0$ the tension of the uniformly covered interface; it is the form appropriate to a perturbation about a pre-existing uniform concentration, and for a nonlinear isotherm the linear theory depends on the isotherm only through $\sigma_0=\sigma(\Gamma_0)$ and $\beta$.

In each fluid the velocity and pressure $p_j$ satisfy
\begin{equation}
  \rho_j\left(\frac{\partial\mathbf u_j}{\partial t}+\mathbf u_j\cdot\nabla\mathbf u_j\right)=-\nabla p_j+\mu_j\nabla^2\mathbf u_j-\rho_j g\,\mathbf e_z,\qquad \nabla\cdot\mathbf u_j=0,\qquad j=l,u.
  \label{eq:ns}
\end{equation}
At the interface the velocity is continuous and the interface moves with the fluid,
\begin{equation}
  \mathbf u\LO=\mathbf u\UP,\qquad \frac{\partial\eta}{\partial t}+u_r\,\frac{\partial\eta}{\partial r}=u_z\qquad\text{at } z=\eta,
  \label{eq:kin}
\end{equation}
and the traction is balanced by the capillary and Marangoni stresses,
\begin{equation}
  \big[(-p\,\mathsf I+2\mu\mathsf D)\cdot\mathbf n\big]_l^{\,u}=\sigma(\nabla_s\cdot\mathbf n)\,\mathbf n-\nabla_s\sigma ,
  \label{eq:stress}
\end{equation}
where $\mathsf D$ is the rate-of-strain tensor and $[\,\cdot\,]_l^{\,u}$ denotes the jump across the interface. The initial conditions are
\begin{equation}
  \Gamma(r,0)=\Gamma_0+\Gamma'(r),\qquad \eta(r,0)=\eta'(r),\qquad \mathbf u_j(r,z,0)=0 ,
  \label{eq:ic}
\end{equation}
with $\Gamma'$ and $\eta'$ arbitrary, and the far-field conditions are $|\mathbf u_j|\to0$ and $p_j\to$ hydrostatic as $z\to\pm\infty$ and as $r\to\infty$. Equations \eqref{eq:surf-full}-\eqref{eq:ic} reduce, mode by mode, to the problem of \citet{Shardt2016oscillatory} when $\eta\equiv0$ is imposed and the upper fluid is removed (\S\ref{sec:analytical}).

\subsection{Non-dimensional form}
\label{sec:nondim}

Let $\lambda$ be the length scale of the initial distribution ($2\pi/k$ for the Bessel mode $J_0(kr)$) and $\varepsilon\Gamma_0$ the amplitude of its variation in concentration, so that $\varepsilon\beta$ is the amplitude of the variation in surface tension. Balancing the Marangoni stress $\varepsilon\beta/\lambda$ against the viscous stress $\mu\LO U/\lambda$ of the lower fluid gives the velocity scale $\varepsilon u^*$ with
\begin{equation}
  u^*=\frac{\beta}{\mu\LO},\qquad t^*=\frac{\lambda}{u^*}=\frac{\lambda\mu\LO}{\beta},
  \label{eq:scales}
\end{equation}
and we set
\begin{equation}
  \tilde r=\frac{r}{\lambda},\quad \tilde z=\frac{z}{\lambda},\quad \tr=\frac{t}{t^*},\quad
  \tilde{\mathbf u}=\frac{\mathbf u}{\varepsilon u^*},\quad
  \tilde p=\frac{p\lambda}{\varepsilon\beta},\quad
  \tilde\Gamma=\frac{\Gamma-\Gamma_0}{\varepsilon\Gamma_0},\quad
  \tilde\eta=\frac{\eta}{\varepsilon\lambda},
  \label{eq:nondim}
\end{equation}
where $\tilde p$ is the departure from the hydrostatic pressure. The interface displacement is scaled with the same amplitude $\varepsilon$ as the concentration, which keeps the kinematic condition free of parameters; an initial displacement of a different amplitude is accommodated by the linearity of the problem. The equations then contain
\begin{equation}
\begin{gathered}
  \Rey=\frac{\rho\LO u^*\lambda}{\mu\LO}=\frac{\rho\LO\beta\lambda}{\mu\LO^2},\qquad
  E=\frac{\beta}{\sigma_0},\qquad
  \Bo=\frac{(\rho\LO-\rho\UP)g\lambda^2}{\sigma_0},\\
  \rho_r=\frac{\rho\UP}{\rho\LO},\qquad
  \mu_r=\frac{\mu\UP}{\mu\LO},
\end{gathered}
  \label{eq:groups}
\end{equation}
together with $\varepsilon$ and the surface P\'eclet number $\Pes=u^*\lambda/D_s$. The Reynolds number is that of \citet{Shardt2016oscillatory}; it compares the viscous time $\rho\LO\lambda^2/\mu\LO$ with the advective time $t^*$. The elasticity number $E$ compares the Marangoni stress with the capillary stress $\sigma_0/\lambda$ that resists deformation of the interface, and $\Bo$ compares gravity with capillarity; both are new, and both enter only through the normal stress balance. A flat interface is recovered for $E\to0$, when the interface is too stiff to be displaced by the Marangoni stress. The Ohnesorge number of capillary-wave theory, $\Oh=\mu\LO/(\rho\LO\sigma_0\lambda/2\pi)^{1/2}$, is not independent: $\Rey=2\pi E/\Oh^2$.

In these variables the surfactant equation and the momentum equations read
\begin{gather}
  \frac{\partial\tilde\Gamma}{\partial\tr}+\varepsilon\,\tilde\nabla_s\cdot(\tilde\Gamma\tilde{\mathbf u}_s)+\tilde\nabla_s\cdot\tilde{\mathbf u}_s+(1+\varepsilon\tilde\Gamma)(\tilde\nabla_s\cdot\mathbf n)(\tilde{\mathbf u}\cdot\mathbf n)=\Pes^{-1}\tilde\nabla_s^2\tilde\Gamma,
  \label{eq:surf-nd}\\
  \Rey\,\hat\rho_j\left(\frac{\partial\tilde{\mathbf u}_j}{\partial\tr}+\varepsilon\,\tilde{\mathbf u}_j\cdot\tilde\nabla\tilde{\mathbf u}_j\right)=-\tilde\nabla\tilde p_j+\hat\mu_j\tilde\nabla^2\tilde{\mathbf u}_j,\qquad \tilde\nabla\cdot\tilde{\mathbf u}_j=0,
  \label{eq:ns-nd}
\end{gather}
with $\hat\rho\LO=\hat\mu\LO=1$, $\hat\rho\UP=\rho_r$ and $\hat\mu\UP=\mu_r$, and the tangential and normal stress conditions at the interface become
\begin{gather}
  \Big[\hat\mu\,\mathbf t\cdot\big(\tilde\nabla\tilde{\mathbf u}+\tilde\nabla\tilde{\mathbf u}^{\rm T}\big)\cdot\mathbf n\Big]_u^{\,l}=-\mathbf t\cdot\tilde\nabla_s\tilde\Gamma ,
  \label{eq:tang-nd}\\
  \Big[\tilde p-\hat\mu\,\mathbf n\cdot\big(\tilde\nabla\tilde{\mathbf u}+\tilde\nabla\tilde{\mathbf u}^{\rm T}\big)\cdot\mathbf n\Big]_u^{\,l}=\frac{1}{E}\Big[(1-\varepsilon E\tilde\Gamma)\,\tilde\kappa+\Bo\,\tilde\eta\Big],
  \label{eq:norm-nd}
\end{gather}
where $\mathbf t$ is the unit tangent in the $(r,z)$ plane, $[\,\cdot\,]_u^{\,l}$ the value in the lower fluid minus that in the upper, and $\tilde\kappa=\varepsilon^{-1}\tilde\nabla_s\cdot\mathbf n$ is the curvature scaled with the amplitude of the displacement: since $\eta=\varepsilon\lambda\tilde\eta$, the geometric curvature $\tilde\nabla_s\cdot\mathbf n$ is of order $\varepsilon$, and $\tilde\kappa=-\tilde\nabla^2\tilde\eta+O(\varepsilon^2)$; the factor $1-\varepsilon E\tilde\Gamma=\sigma/\sigma_0$ follows from \eqref{eos}. Equation \eqref{eq:tang-nd} is equation (2.8) of \citet{Shardt2016oscillatory} supplemented by the stress of the upper fluid, and \eqref{eq:norm-nd} is the condition that the flat-surface problem replaces by $\tilde u_z=0$.

\subsection{Linear problem}

We consider small-amplitude disturbances, $\varepsilon\to0$, and negligible surface diffusion, $\Pes\to\infty$, as in the analytical part of \citet{Shardt2016oscillatory}. The nonlinear terms of \eqref{eq:surf-nd}-\eqref{eq:norm-nd} drop out, the interface conditions are applied at $\tilde z=0$, and the dilatation term of \eqref{eq:surf-nd} is of second order because the base interface is flat. The linear problem is
\begin{gather}
  \frac{\partial\tilde\Gamma}{\partial\tr}+\frac{1}{\tilde r}\frac{\partial(\tilde r\tilde u_{r,s})}{\partial\tilde r}=0,\qquad
  \frac{\partial\tilde\eta}{\partial\tr}=\tilde u_{z,s},
  \label{eq:lin-surf}\\
  \Rey\,\hat\rho_j\frac{\partial\tilde{\mathbf u}_j}{\partial\tr}=-\tilde\nabla\tilde p_j+\hat\mu_j\tilde\nabla^2\tilde{\mathbf u}_j,\qquad \tilde\nabla\cdot\tilde{\mathbf u}_j=0,
  \label{eq:lin-ns}
\end{gather}
with, at $\tilde z=0$,
\begin{gather}
  \tilde u_{r,l}=\tilde u_{r,u}=\tilde u_{r,s},\qquad \tilde u_{z,l}=\tilde u_{z,u}=\tilde u_{z,s},
  \label{eq:lin-cont}\\
  \left(\frac{\partial\tilde u_{r,l}}{\partial\tilde z}+\frac{\partial\tilde u_{z,l}}{\partial\tilde r}\right)-\mu_r\left(\frac{\partial\tilde u_{r,u}}{\partial\tilde z}+\frac{\partial\tilde u_{z,u}}{\partial\tilde r}\right)=-\frac{\partial\tilde\Gamma}{\partial\tilde r},
  \label{eq:lin-tang}\\
  \tilde p\LO-\tilde p\UP-2\frac{\partial\tilde u_{z,l}}{\partial\tilde z}+2\mu_r\frac{\partial\tilde u_{z,u}}{\partial\tilde z}=\frac{1}{E}\left[\Bo\,\tilde\eta-\frac{1}{\tilde r}\frac{\partial}{\partial\tilde r}\left(\tilde r\frac{\partial\tilde\eta}{\partial\tilde r}\right)\right],
  \label{eq:lin-norm}
\end{gather}
the initial conditions $\tilde\Gamma(\tilde r,0)=\tilde\Gamma_0(\tilde r)$, $\tilde\eta(\tilde r,0)=\tilde\eta_0(\tilde r)$, $\tilde{\mathbf u}=0$, and decay of the velocity and pressure at infinity. As in the flat-surface problem the dilatation term of \eqref{eq:lin-surf} must be retained: it is the only way in which the concentration changes. Since the base interface is flat, the perturbation of $\sigma$ enters the normal stress balance only at second order, and within linear theory the surfactant acts on the flow only through the Marangoni term of \eqref{eq:lin-tang}. A clean interface of tension $\sigma_0$ is therefore the natural reference.

\section{Analytical solution}
\label{sec:analytical}

The initial distributions $\tilde\Gamma_0(\tilde r)$ and $\tilde\eta_0(\tilde r)$ are arbitrary, subject only to their possessing Hankel transforms (Fourier transforms in the planar problem). The axisymmetric problem is solved by Hankel transforms in $\tilde r$ and Laplace ($\Lap$) transforms in $\tr$. The Hankel transform of order $\nu$ and its inverse are
\begin{equation}
  \hat f(k)=\mathcal H_\nu\{f\}=\int_0^\infty f(\tilde r)J_\nu(k\tilde r)\,\tilde r\,\mathrm d\tilde r,\qquad
  f(\tilde r)=\int_0^\infty\hat f(k)J_\nu(k\tilde r)\,k\,\mathrm dk ,
  \label{eq:hankel}
\end{equation}
and, following the notation of \citet{Shardt2016oscillatory} with the Fourier transform replaced by the Hankel transform, we write
\begin{equation}
  \hat p(k,\tilde z,\tr)=\mathcal H_0\{\tilde p\},\qquad \breve p(k,\tilde z,s)=\Lap\{\hat p\},
  \label{eq:transforms}
\end{equation}
shown here for the pressure, where $k>0$ is the radial wavenumber and $s$ the complex frequency. The scalars $\tilde p$, $\tilde u_z$, $\tilde\eta$ and $\tilde\Gamma$ are transformed with $\mathcal H_0$ and the radial velocity $\tilde u_r$ with $\mathcal H_1$, because $\partial J_0(kr)/\partial r=-kJ_1(kr)$ and $r^{-1}\partial[rJ_1(kr)]/\partial r=kJ_0(kr)$: a mode $J_0(kr)$ of the pressure, the displacement and the concentration drives a mode $J_1(kr)$ of the radial velocity, and the radial derivatives of \eqref{eq:lin-surf}-\eqref{eq:lin-norm} become multiplications by $\pm k$ or $k^2$ exactly as $\partial/\partial x$ does for the planar modes $\cos kx$ and $\sin kx$.

\subsection{Bulk solutions}

The divergence of \eqref{eq:lin-ns} gives $\tilde\nabla^2\tilde p_j=0$, so that the transformed pressures decay away from the interface as $\mathrm e^{\mp k\tilde z}$. The velocity is described by the Stokes stream function $\tilde\Psi_j$, $\tilde u_{r,j}=-\tilde r^{-1}\partial\tilde\Psi_j/\partial\tilde z$, $\tilde u_{z,j}=\tilde r^{-1}\partial\tilde\Psi_j/\partial\tilde r$, whose single-mode form $\tilde\Psi_j=\tilde rJ_1(k\tilde r)F_j(\tilde z,\tr)$ gives $\tilde u_{r,j}=-F_j'J_1$ and $\tilde u_{z,j}=kF_jJ_0$. The curl of \eqref{eq:lin-ns} reduces to $(\partial_{\tilde z}^2-k^2)(\partial_{\tilde z}^2-k^2-\Rey s\,\hat\rho_j/\hat\mu_j)\breve F_j=0$ for the Laplace transform, because the fluids start from rest, and the solutions bounded as $\tilde z\to\mp\infty$ are
\begin{equation}
\begin{aligned}
  \breve F\LO&=C\LO\,\mathrm e^{k\tilde z}+D\LO\,\mathrm e^{m\LO\tilde z}, &\quad \breve p\LO&=-\Rey\,s\,C\LO\,\mathrm e^{k\tilde z},\\
  \breve F\UP&=C\UP\,\mathrm e^{-k\tilde z}+D\UP\,\mathrm e^{-m\UP\tilde z}, &\quad \breve p\UP&=\Rey\,\rho_r\,s\,C\UP\,\mathrm e^{-k\tilde z},
\end{aligned}
\label{eq:bulk}
\end{equation}
where
\begin{equation}
  m\LO=\sqrt{k^2+\Rey\,s},\qquad m\UP=\sqrt{k^2+\Rey\,s\,\rho_r/\mu_r},\qquad \mathcal R\,m_j>0 ,
  \label{eq:m}
\end{equation}
where $\mathcal R$ and $\mathcal I$ denote real and imaginary parts, to distinguish them from the Reynolds number $\Rey$.
The terms in $C_j$ are irrotational and carry the pressure; those in $D_j$ are the vortical boundary layers, of thickness $|m_j|^{-1}$, which carry no pressure. For the lower fluid alone with $\tilde u_z=0$ at $\tilde z=0$, and with $J_0$ and $J_1$ replaced by $\cos$ and $\sin$, these are equations (3.5)-(3.8) of \citet{Shardt2016oscillatory}.

\subsection{Interface conditions and solution}

The transformed surfactant and kinematic equations \eqref{eq:lin-surf} are
\begin{equation}
  s\Gb-\Gh_0(k)+k\,\breve u_{r,s}=0,\qquad s\eb-\eh_0(k)=\breve u_{z,s},
  \label{eq:transformed-ic}
\end{equation}
with $\breve u_{r,s}=-\breve F'(0)=-(kC\LO+m\LO D\LO)$ and $\breve u_{z,s}=k\breve F(0)=k(C\LO+D\LO)$, and they eliminate $\Gb$ and $\eb$ from the stress conditions. Continuity of velocity then fixes $C_j$ and $D_j$ in terms of the two surface velocities, and the tangential stress condition \eqref{eq:lin-tang}, in which $\partial\tilde u_r/\partial\tilde z+\partial\tilde u_z/\partial\tilde r$ transforms to $-(\breve F''+k^2\breve F)$ and $-\partial\tilde\Gamma/\partial\tilde r$ to $k\Gb$, and the normal stress condition \eqref{eq:lin-norm}, in which the curvature transforms to $k^2\eb$, become two equations for $\breve u_{r,s}$ and $\breve u_{z,s}$,
\begin{equation}
\begin{pmatrix}
Z_t+k^2/s & Z_c\\
Z_c & Z_n+G/s
\end{pmatrix}
\begin{pmatrix}
\breve u_{r,s}\\
\breve u_{z,s}
\end{pmatrix}
=\frac1s
\begin{pmatrix}
k\Gh_0\\
-G\eh_0
\end{pmatrix},
\label{eq:impedance}
\end{equation}
where $G=(\Bo+k^2)/E$ is the dimensionless restoring stiffness of the interface and
\begin{equation}
\begin{gathered}
  Z_t=m\LO+k+\mu_r(m\UP+k),\qquad
  Z_c=m\LO-k-\mu_r(m\UP-k),\\
  Z_n=\frac{m\LO(m\LO+k)+\mu_rm\UP(m\UP+k)}{k}.
\end{gathered}
  \label{eq:Z}
\end{equation}
This form is the clearest statement of the physics. $Z_t$ is the shear resistance of the two fluids to a tangential motion of the interface and $Z_n$ their resistance, viscous and inertial, to a normal one; $k^2/s$ is the elastic stiffness of the surfactant layer, which the tangential motion compresses, and $G/s$ the capillary-gravity stiffness, which the normal motion deforms. The off-diagonal coefficient $Z_c$ is the only coupling between surfactant redistribution and interface deformation: a tangential surface velocity produces a normal stress on the interface, and a normal one a tangential stress, both in proportion to $Z_c$. Its two limits contain most of the results of this paper. For $\Rey\to0$, $m_j-k\simeq\Rey\,s\,\hat\rho_j/(2k\hat\mu_j)$ and
\begin{equation}
  Z_c\simeq\frac{\Rey\,s\,(1-\rho_r)}{2k},
  \label{eq:Zc-stokes}
\end{equation}
so that the coupling vanishes in Stokes flow, is inertial, and has the sign of the density contrast (\S\ref{sec:stokes}). For $\Rey\to\infty$, when both viscous penetration depths $|m\LO|^{-1}$ and $|m\UP|^{-1}$ are small compared with the wavelength, i.e.\ $\Rey|s|\gg k^2$ and $\Rey|s|\rho_r/\mu_r\gg k^2$, $m\LO\simeq(\Rey\,s)^{1/2}$ and $m\UP\simeq(\rho_r/\mu_r)^{1/2}(\Rey\,s)^{1/2}$, and
\begin{equation}
  Z_c\simeq(1-\phi)\,(\Rey\,s)^{1/2},\qquad \phi=(\rho_r\mu_r)^{1/2},
  \label{eq:Zc-large}
\end{equation}
the difference between the shear stresses of the two vortical boundary layers, which changes sign at $\phi=1$ (\S\ref{sec:largeRe}). The coupling vanishes identically, for every $s$, only if $m\UP=m\LO$ and $\mu_r=1$, that is, for
\begin{equation}
  \rho_r=1,\qquad\mu_r=1;
  \label{eq:exact-decoupling}
\end{equation}
for $\phi=1$ with $\rho_r\ne1$ the coupling is suppressed only at leading order in $\Rey^{-1/2}$.

Solving \eqref{eq:impedance} and using \eqref{eq:transformed-ic} gives the Laplace-domain concentration and displacement,
\begin{align}
  \Gb(k,s)&=\frac{\Nc(s)}{s\,\NN(s)}\,\Gh_0(k)-\frac{k^2G\,\QQ(s)}{s^2\,\NN(s)}\,\eh_0(k),
  \label{eq:Gamma-sol}\\
  \eb(k,s)&=\frac{\Nc(s)-kG\Aa+k^2\Cc}{s\,\NN(s)}\,\eh_0(k)-\frac{k^2\QQ(s)}{s^2\,\NN(s)}\,\Gh_0(k),
  \label{eq:eta-sol}
\end{align}
where, after the even powers of $m_j$ have been eliminated,
\begin{equation}
\begin{gathered}
  \Aa=\rho_r(k-m\LO)+(k-m\UP),\qquad
  \B=(k-m\UP)-\rho_r(k-m\LO),\\
  \Cc=k(m\LO+\rho_rm\UP)-(1+\rho_r)m\LO m\UP,\\
  \Nc(s)=\Rey\,s^2\big[(1+\rho_r)\Aa-4\rho_rk\big]+4k^2(1-\mu_r)s\B\\
  \qquad\qquad+\frac{4k^3(1-\mu_r)^2}{\Rey}(k-m\LO)(k-m\UP)+kG\Aa,\\
  \NN(s)=\Nc(s)+k^2\Cc-\frac{k^3G}{\Rey\,s^2}(m\LO-k)(m\UP-k),\\
  \QQ(s)=\frac{2k(1-\mu_r)}{\Rey}(k-m\LO)(k-m\UP)+s\B .
\end{gathered}
\label{eq:N-defs}
\end{equation}
The zeros of $\NN(s)$ are those of the determinant of \eqref{eq:impedance}, $(Z_t+k^2/s)(Z_n+G/s)-Z_c^2$, and the closed form \eqref{eq:Gamma-sol}-\eqref{eq:N-defs} is used below for the asymptotic analysis. The two forms were checked against one another: the transfer functions obtained from \eqref{eq:impedance} and from \eqref{eq:Gamma-sol}-\eqref{eq:eta-sol} agree to round-off for gas-liquid, liquid-liquid and single-fluid parameters at arbitrary complex $s$. $\Nc$ is the dispersion function of a clean interface between two viscous fluids \citep{prosperetti1981motion}, and the remaining two terms of $\NN$ are the Marangoni contribution. The same kernel $k^2\QQ/(s^2\NN)$ transfers an initial concentration to the displacement and, multiplied by $G$, an initial displacement to the concentration: the two problems are reciprocal, as the symmetry of \eqref{eq:impedance} requires. Because every radial derivative in \eqref{eq:lin-surf}-\eqref{eq:lin-norm} acts on a Bessel mode exactly as it acts on a planar mode, \eqref{eq:impedance}-\eqref{eq:N-defs} are also the transfer functions of the planar problem, with $\mathcal H_0$ and $\mathcal H_1$ replaced by the Fourier transform and $J_0(kr)$, $J_1(kr)$ by $\cos kx$, $\sin kx$; what distinguishes the two geometries is the mode shape and the way the modes combine in \eqref{eq:hankel}.

Three limits check the solution. For a flat interface, $E\to0$ so that $G\to\infty$, the terms in $G$ dominate $\Nc$ and $\NN$, the displacement vanishes, and
\begin{equation}
  \Gb(k,s)\to\frac{\Gh_0(k)}{s+\dfrac{k^2}{\Rey\,s}\,\dfrac{(m\LO-k)(m\UP-k)}{\rho_r(m\LO-k)+(m\UP-k)}},
  \label{eq:flat-limit}
\end{equation}
in which the shear stresses of the two boundary layers add: for $\Rey\to\infty$ the fraction tends to $(\Rey\,s)^{1/2}/(1+\phi)$, so that the upper fluid increases the resistance to the Marangoni flow by the factor $1+\phi$. When the upper fluid is removed the fraction reduces to $m\LO-k$ and
\begin{equation}
  \Gb(k,s)\to\frac{\Gh_0(k)}{s+\dfrac{k^2}{m\LO+k}}=\frac{\Gh_0(k)}{s+\dfrac{k^3}{\Rey\,s}\Big(\sqrt{1+\dfrac{\Rey\,s}{k^2}}-1\Big)},
  \label{eq:shardt-limit}
\end{equation}
which is equation (3.13) of \citet{Shardt2016oscillatory}: their planar solution holds, mode by mode, for the axisymmetric flat-surface problem. The clean interface is a singular limit of the present scaling, because both $\Rey$ and $E$ are proportional to $\beta$; it is taken in dimensional variables, $\beta\to0$ at fixed $\rho_j$, $\mu_j$, $\sigma_0$ and $\lambda$, in which the Marangoni terms of $\NN$, proportional to $\beta$, vanish while $\Nc$ remains, and the poles of $\eb$ are then the roots of $\Nc=0$, the two-fluid dispersion relation of \citet{prosperetti1981motion}. For an inextensible interface, $E\to\infty$ with $\sigma_0$ fixed, the Marangoni terms must vanish by themselves, $s^2\Cc=kG(m\LO-k)(m\UP-k)/\Rey$, which is equivalent to $\breve u_{r,s}=0$. When the upper fluid is removed, $\rho_r,\mu_r\to0$ at fixed $\rho_r/\mu_r$, every term of $\NN$, $\Nc$ and $\QQ$ carries the factor $k-m\UP$, which cancels, and \eqref{eq:Gamma-sol}-\eqref{eq:eta-sol} reduce to the single-fluid problem with a deformable free surface; its dispersion function is the Levich-Lucassen dispersion relation for capillary-gravity waves on a liquid covered by an insoluble elastic film \citep{levich1962physicochemical,lucassen1970properties}.

\subsection{Inversion}
\label{sec:inversion}

The square roots \eqref{eq:m} have branch points at
\begin{equation}
  s_{b,l}=-\frac{k^2}{\Rey},\qquad s_{b,u}=-\frac{k^2\mu_r}{\Rey\,\rho_r},
  \label{eq:branchpoints}
\end{equation}
which are the inverse viscous diffusion times of the two fluids, and we take the physical sheet $\mathcal R\,m_j>0$, on which the vortical parts of \eqref{eq:bulk} decay away from the interface, with the cuts along the negative real axis. The Bromwich integral is closed in the left half-plane (Appendix~\ref{sec:appA}) and the time-domain solution is a sum over the poles $s_n$ of \eqref{eq:Gamma-sol}, the zeros of $\NN$ on the physical sheet, and an integral along the cut:
\begin{equation}
  \Gh(k,\tr)=\sum_n\mathrm{Res}_{s_n}\big[\Gb\big]\,\mathrm e^{s_n\tr}-\frac1\pi\int_{x_{\min}}^\infty\mathrm e^{-x\tr}\,\mathcal I\big[\Gb^+(k,-x)\big]\,\mathrm dx,
  \label{eq:inversion}
\end{equation}
where $x_{\min}=\min(-s_{b,l},-s_{b,u})$, $\Gb^+(k,-x)$ is $\Gb$ evaluated on the upper bank of the cut, $s=-x+\mathrm i0$, and the same formula holds for $\eh$. The poles are the normal modes of the interface. Every pole of a transfer function is a zero of $\NN$, but a zero of $\NN$ need not be a pole of every transfer function: a numerator can vanish at it. In the decoupled case \eqref{eq:exact-decoupling} the determinant of \eqref{eq:impedance} factorizes into $(Z_t+k^2/s)(Z_n+G/s)$, the concentration has only the poles of the first factor and the displacement only those of the second, and the cross transfer functions vanish. The poles are real or occur in complex-conjugate pairs; a complex pole represents a damped oscillation of frequency $\mathcal I\,s_n$ and decay rate $-\mathcal R\,s_n$. The cut integral represents the diffusion of vorticity into the bulk of each fluid; at late times it decays as $\mathrm e^{-x_{\min}\tr}\tr^{-3/2}$ (Appendix~\ref{sec:appA}), which is algebraic only in the limit $x_{\min}\to0$ of large $\Rey$. The discrete modes alone do not satisfy the initial conditions: the residues differ from the weights that a normal-mode expansion would assign to the same exponentials, because the contribution from continuous spectrum must be included.

\section{Solution for a single mode}
\label{sec:single}

By linearity, the solution for any axisymmetric initial distribution is the Hankel integral \eqref{eq:hankel} of the solutions for its Bessel modes $J_0(kr)$, and the modes of smallest $k$, which decay slowest, determine the late-time behaviour; the Cauchy-Poisson problem of a localized surfactant deposit is this integral. We therefore consider the single mode
\begin{equation}
  \tilde\Gamma_0(\tilde r)=J_0(2\pi\tilde r),\qquad \tilde\eta_0(\tilde r)=0,
  \label{eq:single-ic}
\end{equation}
for which $\Gh_0(k)=\delta(k-2\pi)/2\pi$ and only $k=2\pi$ contributes. Writing $\theta_\Gamma(\tr)$ and $\theta_\eta(\tr)$ for the time-dependent amplitudes,
\begin{equation}
  \tilde\Gamma(\tilde r,\tr)=\theta_\Gamma(\tr)J_0(2\pi\tilde r),\qquad \tilde\eta(\tilde r,\tr)=\theta_\eta(\tr)J_0(2\pi\tilde r),
  \label{eq:theta-def}
\end{equation}
where $\theta_\Gamma$ and $\theta_\eta$ are the inverse Laplace transforms of $\Nc/(s\NN)$ and $-k^2\QQ/(s^2\NN)$ at $k=2\pi$, with $\theta_\Gamma(0)=1$ and $\theta_\eta(0)=0$. Since $\tilde\Gamma$ is the perturbation of the concentration from $\Gamma_0$, $\theta_\Gamma$ takes both signs: negative values mean that the concentration on the axis is below $\Gamma_0$, not that it is negative. The concentration is highest on the axis, where $J_0=1$, and the mode is a central patch surrounded by rings of alternating sign. The problem depends on $\Rey$, $E$, $\Bo$, $\rho_r$ and $\mu_r$. Two combinations of the fluid properties recur: the ratio
\begin{equation}
  \phi=\left(\frac{\rho\UP\mu\UP}{\rho\LO\mu\LO}\right)^{1/2}=(\rho_r\mu_r)^{1/2},
  \label{eq:phi}
\end{equation}
which compares the shear stresses that the two vortical boundary layers exert on the interface, and the inviscid frequency of the capillary-gravity wave in units of $1/t^*$,
\begin{equation}
  \tilde\omega_0=\left[\frac{k(k^2+\Bo)}{(1+\rho_r)\,E\,\Rey}\right]^{1/2}=\left[\frac{(2\pi)^3(1+\Bo/4\pi^2)}{(1+\rho_r)\,E\,\Rey}\right]^{1/2}.
  \label{eq:omega0}
\end{equation}
The dependence on $E$ and $\Bo$ is mainly through $\tilde\omega_0$. Unless stated otherwise the results below are for $\Bo=0$; gravity only increases the stiffness of the interface. Since the transfer functions are those of the planar problem, everything in this section, and in particular the comparison with the flat-surface results of \citet{Shardt2016oscillatory}, holds equally for the planar mode $\cos kx$.

\subsection{Poles of the Laplace-domain solution}
\label{sec:poles}

\begin{figure}
  \centering
  \includegraphics[width=\textwidth]{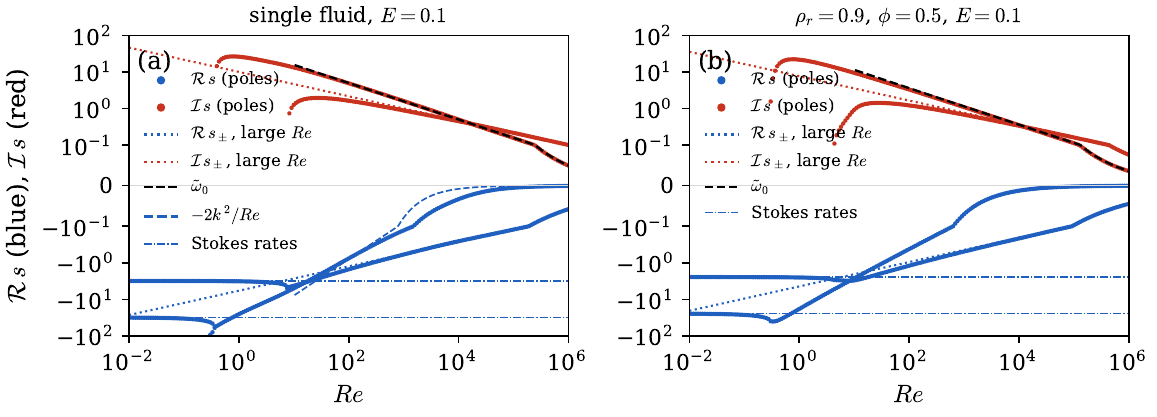}
  \caption{Poles of the Laplace-domain solution \eqref{eq:Gamma-sol} for the single mode $k=2\pi$, for $E=0.1$ and $\Bo=0$: (a) a single fluid with a free surface; (b) two liquids with $\rho_r=0.9$ and $\phi=0.5$. Blue dots show the real parts $\mathcal R\,s$ of the poles and red dots their imaginary parts $\mathcal I\,s$ (only one of each conjugate pair is shown). Blue and red dotted lines: real and imaginary parts of the large-$\Rey$ Marangoni poles \eqref{eq:marangoni-poles}. Black dashed line: the inviscid wave frequency $\tilde\omega_0$ of \eqref{eq:omega0}. Blue dashed line, in (a): the clean viscous damping $-2k^2/\Rey$. Blue dash-dotted lines: the Stokes-limit decay rates \eqref{eq:stokes-rates}. The vertical scale is logarithmic above $|s|=0.1$ and linear below.}
  \label{fig:poles}
\end{figure}

Figure~\ref{fig:poles} shows the poles of \eqref{eq:Gamma-sol} as functions of $\Rey$ for $E=0.1$, for a single fluid with a free surface and for a pair of liquids. Compared with figure~2 of \citet{Shardt2016oscillatory} there are twice as many poles: the two Marangoni poles of the flat-surface problem are joined by the two poles of the capillary-gravity wave, and the dynamics is classified by the positions of both pairs.

For $\Rey\to0$ there are two negative real poles, at the Stokes-limit relaxation rates of the surfactant and of the interface (\S\ref{sec:stokes}),
\begin{equation}
  s_\Gamma=-\frac{k}{2(1+\mu_r)}=-\frac{\pi}{1+\mu_r},\qquad
  s_\eta=-\frac{G}{2k(1+\mu_r)}=-\frac{\pi(1+\Bo/4\pi^2)}{E(1+\mu_r)}.
  \label{eq:stokes-rates}
\end{equation}
As $\Rey$ increases a third real pole enters from the branch point $-k^2/\Rey$, and at $\Rey\simeq0.4$ it merges with $s_\eta$ into a complex pair: the interface mode changes from overdamped to underdamped, exactly as the capillary wave on a clean viscous liquid does \citep{prosperetti1976}, and its frequency then approaches $\tilde\omega_0\propto\Rey^{-1/2}$ from below while its damping approaches the clean value $-2k^2/\Rey$. The surfactant pole $s_\Gamma$ moves slowly to more negative values, $s_\Gamma\simeq-\{k/[2(1+\mu_r)]\}[1+(1+\rho_r)\Rey/(8k(1+\mu_r)^2)]$ to first order in $\Rey$, meets a second real pole, and the two become a complex pair at $\Rey/2\pi\simeq32/27$, the critical Reynolds number of \citet{Shardt2016oscillatory}; the deformability of the interface changes this threshold by only a few per cent at $E=0.1$ (Table~\ref{tab:critical}). The upper fluid moves it in either direction: its inertia, which adds to the Reynolds number, lowers the threshold ($\phi=0.5$), while its viscosity, which adds to the resistance, raises it ($\phi=1$, $2$); for the gas-liquid interface it is indistinguishable from the free surface. Beyond this value the Marangoni pair approaches the large-$\Rey$ asymptote
\begin{equation}
  s_\pm=\frac{2\pi}{(1+\phi)^{2/3}}\left(\frac{\Rey}{2\pi}\right)^{-1/3}\left(-\frac12\pm\frac{\mathrm i\sqrt3}{2}\right),
  \label{eq:marangoni-poles}
\end{equation}
the poles (3.17) of \citet{Shardt2016oscillatory} with their modulus reduced by $(1+\phi)^{-2/3}$ (\S\ref{sec:largeRe}).

\begin{figure}
  \centering
  \includegraphics[width=\textwidth]{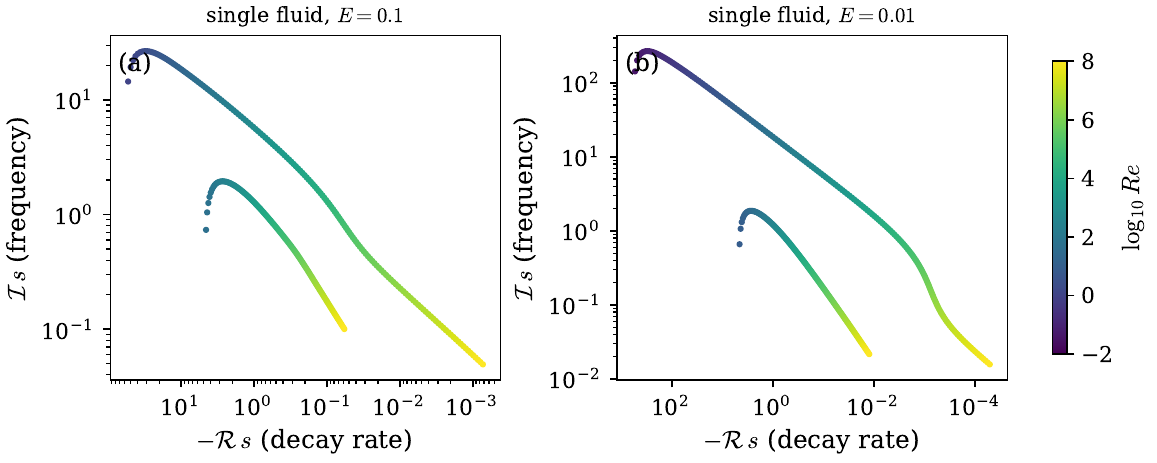}
  \caption{Trajectories of the complex poles of \eqref{eq:Gamma-sol} in the plane of the decay rate $-\mathcal R\,s$ and the frequency $\mathcal I\,s$, coloured by $\log_{10}\Rey$, for a single fluid with a free surface: (a) $E=0.1$, (b) $E=0.01$. The upper branch is the capillary-gravity wave and the lower branch the Marangoni pair; the two approach near $q=1$, equation \eqref{eq:crossing}, without coalescing.}
  \label{fig:trajectories}
\end{figure}

The two pairs do not evolve independently. The wave frequency decreases as $\Rey^{-1/2}$ and the Marangoni frequency as $\Rey^{-1/3}$, so their time scales match, $|s_\pm|=\tilde\omega_0$, at
\begin{equation}
  E^3\Rey=\frac{2\pi(1+\phi)^4(1+\Bo/4\pi^2)^3}{(1+\rho_r)^3}.
  \label{eq:crossing}
\end{equation}
For $E=0.1$ and a single fluid this is $\Rey\simeq6.3\times10^3$. It is convenient to measure the distance from this condition by $q=(|s_\pm|/\tilde\omega_0)^{3/2}$, which equals $[E^3\Rey(1+\rho_r)^3/2\pi(1+\phi)^4(1+\Bo/4\pi^2)^3]^{1/4}$, so that \eqref{eq:crossing} is $q=1$. Near $q=1$ the two branches interact: the damping of the wave rises well above its clean value and the Marangoni poles leave their asymptote. This interaction of the longitudinal and transverse branches, and the resonant damping that it produces, are known from the normal-mode analyses of \citet{lucassen1970properties} and \citet{Rajan2020resonance}; it is not a coalescence of poles. Figure~\ref{fig:trajectories} shows the poles in the complex plane: the wave and Marangoni pairs approach one another near $q=1$ but remain distinct, their closest separation along the sweep of figure~\ref{fig:poles}(a) being $44\,\%$ of their modulus, and the conditions $\NN=\NN'=0$ for a double root are not met. Several characteristic values of $q$ describe the interaction, and they should not be confused. Time scales match at $q=1$; the excitation of the wave by the surfactant is largest at $q=1/\sqrt2$ when $\Rey$ is varied at fixed elasticity, and at $q=1$ when the elasticity is varied at fixed $\Oh$; and the damping of the wave is largest at $q=\sqrt2$ (\S\ref{sec:largeRe}). For $E\ll1$ these lie at very large $\Rey$, and for $E=O(1)$ at $\Rey=O(1)$, where the modes are still overdamped. For the two liquids of figure~\ref{fig:poles}(b) the same sequence occurs with the rates \eqref{eq:stokes-rates} reduced by $1+\mu_r$ and the asymptote \eqref{eq:marangoni-poles} reduced by $(1+\phi)^{2/3}$.

\begin{table}
  \begin{center}
  \def~{\hphantom{0}}
  \begin{tabular}{lcccc}
      fluids & $\rho_r$ & $\mu_r$ & $\phi$ & $\Rey_c/2\pi$ \\[3pt]
      single fluid, free surface & 0 & 0 & 0 & 1.2 \\
      liquid-liquid & 0.9 & 0.28 & 0.5 & 0.66 \\
      liquid-liquid & 0.9 & 1.1 & 1 & 2.7 \\
      liquid-liquid & 0.9 & 4.4 & 2 & 9.5 \\
      gas-liquid & $1.2\times10^{-3}$ & $1.8\times10^{-2}$ & $4.6\times10^{-3}$ & 1.3 \\
  \end{tabular}
  \caption{Critical Reynolds number $\Rey_c$ above which the Marangoni poles are complex, for $E=0.1$ and $\Bo=0$. For a flat surface bounding a single fluid, $\Rey_c/2\pi=32/27=1.19$ \citep{Shardt2016oscillatory}.}
  \label{tab:critical}
  \end{center}
\end{table}

\subsection{Stokes flow limit}
\label{sec:stokes}

For $\Rey\to0$ the square roots \eqref{eq:m} tend to $k$, every term of \eqref{eq:N-defs} is of order $\Rey$, and the leading-order solution is
\begin{equation}
  \Gb=\frac{\Gh_0(k)}{s+\dfrac{k}{2(1+\mu_r)}},\qquad
  \eb=\frac{\eh_0(k)}{s+\dfrac{G}{2k(1+\mu_r)}}+O(\Rey).
  \label{eq:stokes-sol}
\end{equation}
The surfactant decays exponentially, as on a flat surface \citep{Shardt2016oscillatory,Bickel2022exact}, at the rate $k/2$ divided by $1+\mu_r$: the Marangoni stress is now balanced by the shear of both fluids. The interface relaxes independently at the overdamped capillary-gravity rate $G/2k(1+\mu_r)$, and the cross terms of \eqref{eq:Gamma-sol}-\eqref{eq:eta-sol} vanish, because $\QQ=O(\Rey^2)$ while $\NN=O(\Rey)$. The surfactant therefore does not displace the interface in Stokes flow. The reason is seen in the flow itself. The time-domain solution for the single mode \eqref{eq:single-ic} is
\begin{equation}
\begin{aligned}
  \theta_\Gamma&=\exp\!\left(-\frac{\pi\tr}{1+\mu_r}\right),\qquad \theta_\eta=0,\\
  \tilde u_{r,l}&=\frac{J_1(2\pi\tilde r)}{2(1+\mu_r)}\,(1+2\pi\tilde z)\,\mathrm e^{2\pi\tilde z}\,\theta_\Gamma, &
  \tilde u_{z,l}&=-\frac{J_0(2\pi\tilde r)}{2(1+\mu_r)}\,2\pi\tilde z\,\mathrm e^{2\pi\tilde z}\,\theta_\Gamma,\\
  \tilde u_{r,u}&=\frac{J_1(2\pi\tilde r)}{2(1+\mu_r)}\,(1-2\pi\tilde z)\,\mathrm e^{-2\pi\tilde z}\,\theta_\Gamma, &
  \tilde u_{z,u}&=-\frac{J_0(2\pi\tilde r)}{2(1+\mu_r)}\,2\pi\tilde z\,\mathrm e^{-2\pi\tilde z}\,\theta_\Gamma,\\
  \tilde p\LO&=-\frac{2\pi J_0(2\pi\tilde r)}{1+\mu_r}\,\mathrm e^{2\pi\tilde z}\,\theta_\Gamma, &
  \tilde p\UP&=-\frac{2\pi\mu_rJ_0(2\pi\tilde r)}{1+\mu_r}\,\mathrm e^{-2\pi\tilde z}\,\theta_\Gamma,
\end{aligned}
\label{eq:stokes-field}
\end{equation}
which for $\mu_r=0$ is the axisymmetric counterpart of the solution (3.30) of \citet{Shardt2016oscillatory}: the surface flow diverges from the axis, where the concentration is highest, fluid rises beneath it, and the toroidal vortices so formed, stretched vertically and reflected in the upper fluid, decay without changing shape. On the interface the pressure of each fluid is $-2\pi\hat\mu_jJ_0(2\pi\tilde r)\,\theta_\Gamma/(1+\mu_r)$ and the viscous normal stress $2\hat\mu_j\partial\tilde u_{z,j}/\partial\tilde z$ is exactly equal to it, so that their contributions to the normal traction cancel and the traction $-\tilde p_j+2\hat\mu_j\partial\tilde v_j/\partial\yr$ vanishes identically in each fluid. A Marangoni stress on a flat interface in Stokes flow produces no normal force on it, for any viscosity ratio, and the flat-surface problem and the deformable-interface problem coincide in this limit. The quantity that the Marangoni flow would exert on a rigid flat surface, the pressure $-k\Gb$ of \citet{Shardt2016oscillatory}, is not what deforms a free interface.

The coupling is an inertial effect. Expanding to first order in $\Rey$ (Appendix~\ref{sec:appB}),
\begin{equation}
  \eb=-\frac{\Rey\,(1-\rho_r)}{8k^2(1+\mu_r)^2}\;\frac{s}{(s+a)(s+b)}\,\Gh_0(k),\qquad a=\frac{k}{2(1+\mu_r)},\quad b=\frac{G}{2k(1+\mu_r)},
  \label{eq:stokes-ORe}
\end{equation}
so that, for the single mode, $\theta_\eta=-[\Rey(1-\rho_r)/8k^2(1+\mu_r)^2]\,(b\,\mathrm e^{-b\tr}-a\,\mathrm e^{-a\tr})/(b-a)$ for $\tr\gg\Rey/k^2$. The displacement is of order $\Rey$, is directed downward beneath a maximum of concentration when the lower fluid is the denser, and its sign is set by the density contrast alone: in the Stokes regime the normal force is the inertia of the Marangoni flow, which is unbalanced only if the two fluids have different densities. The concentration is unaffected by the deformation at this order, $\Gb=\Gh_0[1/(s+a)+\Rey(1+\rho_r)s/8k(1+\mu_r)^2(s+a)^2]$, the correction being the inertial shift of the real pole noted in \S\ref{sec:poles}.

\subsection{Large Reynolds number}
\label{sec:largeRe}

For $\Rey\to\infty$ the boundary layers are thin, $m\LO\simeq L$ and $m\UP\simeq(\rho_r/\mu_r)^{1/2}L$ with $L=(\Rey\,s)^{1/2}\gg k$, and the terms of \eqref{eq:N-defs} are ordered in powers of $L$. Keeping the leading terms (Appendix~\ref{sec:appB}),
\begin{equation}
\begin{gathered}
  \NN\simeq-\Big(\frac{\rho_r}{\mu_r}\Big)^{1/2}(1+\rho_r)\,\Rey\,(s^2+\tilde\omega_0^2)\left[(1+\phi)\sqrt{\Rey\,s}+\frac{k^2}{s}\right],\\
  \Nc\simeq-\Big(\frac{\rho_r}{\mu_r}\Big)^{1/2}(1+\rho_r)\,\Rey\,(s^2+\tilde\omega_0^2)\,(1+\phi)\sqrt{\Rey\,s},
\end{gathered}
  \label{eq:largeRe-N}
\end{equation}
and the dispersion function factorizes into the inviscid wave, $s^2+\tilde\omega_0^2=0$, and the Marangoni factor $(1+\phi)s^{3/2}=-k^2\Rey^{-1/2}$, whose roots are \eqref{eq:marangoni-poles}. The wave factor cancels between $\Nc$ and $\NN$, and the surfactant transfer function reduces to
\begin{equation}
  \frac{\Nc}{s\NN}\simeq\breve\theta(s)=\frac{1}{s+\dfrac{k^2}{(1+\phi)\sqrt{\Rey\,s}}},
  \label{eq:theta-largeRe}
\end{equation}
which is equation (3.16) of \citet{Shardt2016oscillatory} with $\Rey$ replaced by $(1+\phi)^2\Rey$. The inversion of \eqref{eq:theta-largeRe} is therefore their equation (3.23) in the rescaled time
\begin{equation}
  \tau=\frac{(2\pi)^{4/3}}{(1+\phi)^{2/3}\Rey^{1/3}}\,\tr ,\qquad
  \theta(\tau)=\frac13\sum_{n=0,\pm}\exp(b_n^2\tau)\,\erfc(-b_n\sqrt\tau),
  \label{eq:theta-shardt}
\end{equation}
with $b_0=-1$ and $b_\pm=(1\pm\mathrm i\sqrt3)/2$, together with its early- and late-time forms
\begin{equation}
  \theta\simeq1-\frac{4}{3\sqrt\pi}\,\tau^{3/2}\quad(\tau\to0),\qquad
  \theta\simeq-\frac{1}{2\sqrt\pi}\,\tau^{-3/2}\quad(\tau\to\infty).
  \label{eq:theta-asymptotes}
\end{equation}
To leading order at large $\Rey$, then, neither the deformability of the interface nor the elasticity number affects the relaxation of the surfactant, and the upper fluid enters only through $\phi$: the two vortical boundary layers exert shear stresses in the ratio $\phi$, and the surfactant decay time scale $t^*\Rey^{1/3}$ of \citet{Shardt2016oscillatory} becomes
\begin{equation}
  t_\Gamma^*=(1+\phi)^{2/3}\,t^*\Rey^{1/3}=(1+\phi)^{2/3}\left(\frac{\rho\LO\lambda^4\mu\LO}{\beta^2}\right)^{1/3}.
  \label{eq:tGamma}
\end{equation}
In the same approximation the displacement initiated by the surfactant is
\begin{equation}
  -\frac{k^2\QQ}{s^2\NN}\simeq-\frac{k^2\chi}{(1+\rho_r)\Rey}\;\frac{1}{s\,(s^2+\tilde\omega_0^2)\Big[1+\dfrac{k^2}{(1+\phi)\Rey^{1/2}s^{3/2}}\Big]},\qquad
  \chi=\frac{1-\phi}{1+\phi},
  \label{eq:eta-largeRe}
\end{equation}
whose short-time limit, $\theta_\eta\simeq-k^2\chi\,\tr^2/2(1+\rho_r)\Rey$, shows the interface accelerating from rest under the inertia of the Marangoni flow, downward beneath a concentration maximum when $\phi<1$ and upward when $\phi>1$, and whose poles $\pm\mathrm i\tilde\omega_0$ show that what is initiated is the capillary-gravity wave. The wave is a correction of relative order $k^2/\Rey\tilde\omega_0^2=(1+\rho_r)E/2\pi(1+\Bo/4\pi^2)$ to the concentration, which is why it cancels at leading order in \eqref{eq:theta-largeRe}; but it is the least damped discrete mode, and it dominates the intermediate times; whether it also dominates at the latest times depends on its damping relative to the decay of the continuous spectrum (\S\ref{sec:decay}).

The next order in $L^{-1}$ gives the damping of the wave and its interaction with the Marangoni poles. The expansion is that of the dispersion relation for weak viscosity \citep{Rajan2020resonance}, and we quote the result in the present variables (Appendix~\ref{sec:appB}). With $q=(|s_\pm|/\tilde\omega_0)^{3/2}$, which by \eqref{eq:crossing} equals $[E^3\Rey(1+\rho_r)^3/2\pi(1+\phi)^4(1+\Bo/4\pi^2)^3]^{1/4}$, the decay rate of the wave is
\begin{equation}
\begin{gathered}
  \zeta=\zeta_{\rm cl}+\Delta\zeta_M\,\frac{q^2}{1-\sqrt2q+q^2},\\
  \zeta_{\rm cl}=\frac{\sqrt2\,k\,\phi}{(1+\rho_r)(1+\phi)}\Big(\frac{\tilde\omega_0}{\Rey}\Big)^{1/2},\qquad
  \Delta\zeta_M=\frac{k\,\chi^2(1+\phi)}{2\sqrt2(1+\rho_r)}\Big(\frac{\tilde\omega_0}{\Rey}\Big)^{1/2}.
\end{gathered}
  \label{eq:damping}
\end{equation}
The clean damping $\zeta_{\rm cl}$ is the boundary-layer dissipation at a liquid-liquid interface, which vanishes for a free surface, where the damping is the smaller bulk value $2k^2/\Rey$. The Marangoni contribution rises from zero at $q\to0$ to $2\Delta\zeta_M$ at $q=\sqrt2$, so that the largest damping is $\zeta_{\rm cl}+2\Delta\zeta_M$, and it returns to $\Delta\zeta_M$ as $q\to\infty$, where $\zeta_{\rm cl}+\Delta\zeta_M$ is the damping of a wave on an inextensible interface. The maximum is therefore twice the inextensible value only for a free surface, $\zeta_{\rm cl}=0$; it lies at $E^3\Rey$ four times the value \eqref{eq:crossing}. This resonant damping is the result of \citet{Rajan2020resonance} in the present variables.

\begin{figure}
  \centering
  \includegraphics[width=.6\textwidth]{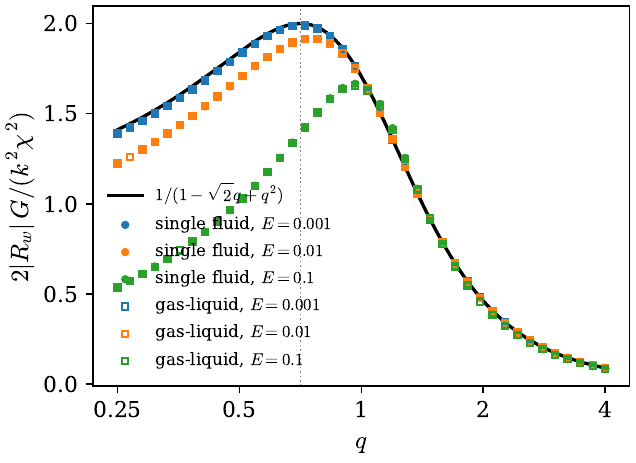}
  \caption{Amplitude of the wave-locked oscillation of the concentration, $2|R_w|$, normalised by $k^2\chi^2/G$, against $q$: exact residues for a single fluid (filled circles) and a gas-liquid interface (open squares) at $E=10^{-3}$ (blue), $10^{-2}$ (orange) and $10^{-1}$ (green), and the boundary-layer result \eqref{eq:residue} (black line). The dotted line marks the maximum at $q=1/\sqrt2$.}
  \label{fig:residue}
\end{figure}

What the normal-mode analysis does not give is how strongly a surfactant disturbance excites the wave. The amplitude of the wave-locked oscillation of the concentration is $2|R_w|$, where $R_w$ is the residue of $\Nc/(s\NN)$ at the wave pole. Evaluated with the boundary-layer forms of $\Nc$ and $\NN$, it is
\begin{equation}
  2|R_w|\simeq\frac{k^2\chi^2}{G}\,\frac{1}{1-\sqrt2q+q^2}
  =\frac{E\chi^2}{1+\Bo/k^2}\,\frac{1}{1-\sqrt2q+q^2},
  \label{eq:residue}
\end{equation}
Its maximum depends on what is varied. At fixed $E$ and $\Bo$, varying $\Rey$, it is largest at $q=1/\sqrt2$, where it equals $2E\chi^2/(1+\Bo/k^2)$, at a quarter of the value of $E^3\Rey$ of \eqref{eq:crossing}. At fixed $\Oh$ and fluid properties, varying the elasticity, $\Rey=2\pi E/\Oh^2$ makes $q\propto E$, so that $2|R_w|\propto q/(1-\sqrt2q+q^2)$, which is largest at $q=1$. The Marangoni part of the damping, normalised by $\Delta\zeta_M$, is largest at $q=\sqrt2$. Figure~\ref{fig:residue} compares \eqref{eq:residue} with the residues computed from the exact dispersion function: for $E=10^{-3}$ the two agree to within $0.1\,\%$ over $0.25<q<4$, for single-fluid and gas-liquid interfaces alike, and the departures at $E=0.1$, confined to $q<1$, are the finite-elasticity corrections. The excitation and the Marangoni part of the damping share the factor $(1-\sqrt2q+q^2)^{-1}$: the same coupling $Z_c$ through which the surfactant drives the wave also removes energy from it, but the excitation at fixed elasticity favours slower Marangoni relaxation ($q<1$) and the damping faster ($q>1$). Both are proportional to $\chi^2$, and both vanish at leading order for $\phi=1$, when $\rho\UP\mu\UP=\rho\LO\mu\LO$; the exact decoupling requires \eqref{eq:exact-decoupling}.

\subsection{Temporal regimes}
\label{sec:decay}

\begin{figure}
  \centering
  \includegraphics[width=\textwidth]{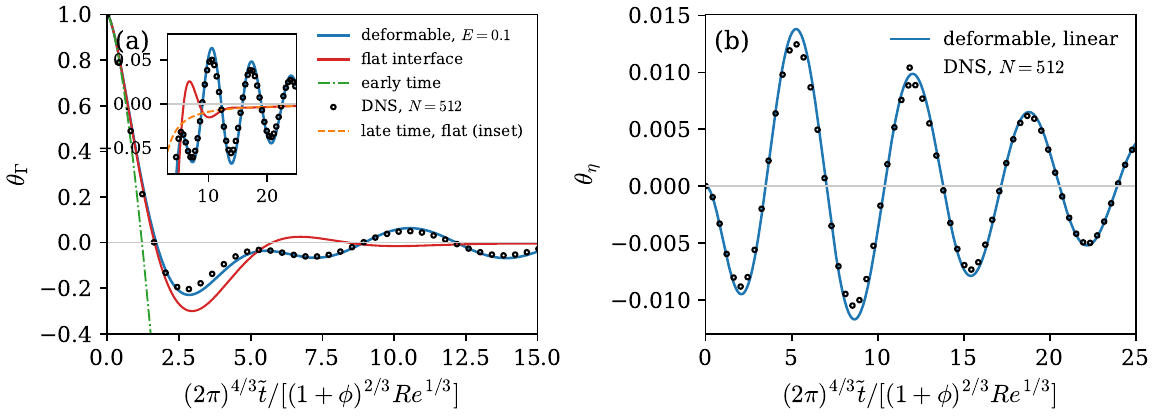}
  \caption{(a) Evolution of the amplitude $\theta_\Gamma$ of the surfactant perturbation $J_0(kr)$ about $\Gamma_0$, which can be negative, on a deformable gas-liquid interface at $\Rey=10^4$, $E=0.1$ and $\Bo=0$: linear solution (blue), flat-surface solution \eqref{eq:theta-shardt} (red), early-time law \eqref{eq:theta-asymptotes} (green dash-dotted) and DNS at $N=512$ (black circles). The inset shows the later behaviour, with the algebraic decay \eqref{eq:theta-asymptotes} of the flat-surface solution as the orange dashed line. (b) The displacement $\theta_\eta$ of the interface initiated by the same surfactant distribution: linear solution (blue) and DNS (black circles). The time is scaled as in \eqref{eq:theta-shardt}.}
  \label{fig:decay}
\end{figure}

Figure~\ref{fig:decay}(a) shows $\theta_\Gamma(\tr)$ for a gas-liquid interface at $\Rey=10^4$ and $E=0.1$, a little beyond the time-scale matching \eqref{eq:crossing}, for which the Marangoni poles are $-0.26\pm0.45\mathrm i$ and the wave poles $-0.033\pm0.50\mathrm i$, and compares it with the flat-surface solution \eqref{eq:theta-shardt} and with the DNS. The deformable and flat surfaces coincide at early times, as \eqref{eq:theta-largeRe} requires, through the first zero at $\tau=1.65$, and part at the first minimum, which is shallower on the deformable surface. On the flat surface the amplitude changes sign twice more and then decays monotonically; on the deformable surface it keeps oscillating at the wave frequency, with a slowly decaying envelope. Figure~\ref{fig:decay}(b) shows the displacement that the surfactant has initiated: it starts as $-\tr^2$, in accordance with \eqref{eq:eta-largeRe}, and settles into the capillary-gravity wave. The surfactant is carried by this wave.

\begin{figure}
  \centering
  \includegraphics[width=\textwidth]{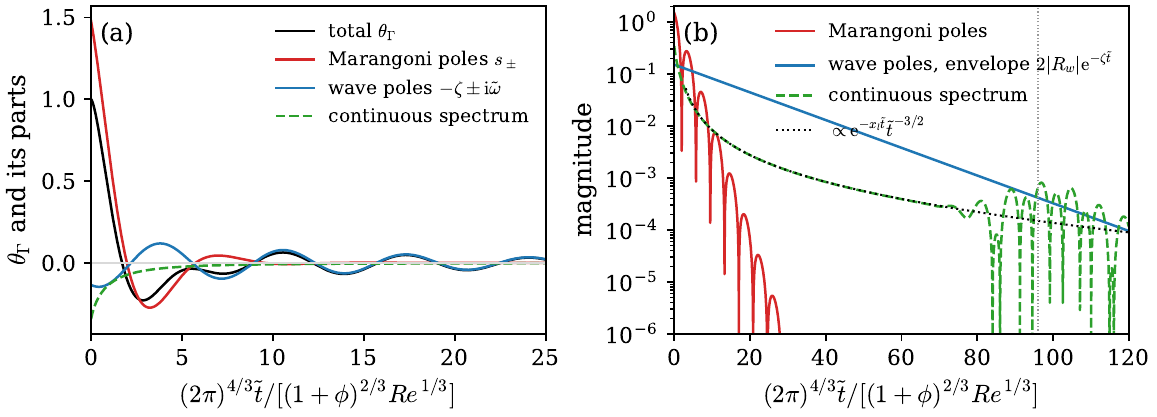}
  \caption{The solution of figure~\ref{fig:decay}(a) separated into its spectral contributions: the residues at the Marangoni poles $s_\pm$ (red), the residues at the wave poles $-\zeta\pm\mathrm i\tilde\omega$ (blue) and the integral along the branch cut, the continuous spectrum (green dashed); the total is black. (a) The contributions; (b) their magnitudes on a logarithmic scale, the wave contribution being represented by its envelope $2|R_w|\mathrm e^{-\zeta\tr}$, with the late-time form $\mathrm e^{-x_l\tr}\tr^{-3/2}$ of the continuous spectrum (black dotted), where $x_l=k^2/\Rey$.}
  \label{fig:decomposition}
\end{figure}

The inversion \eqref{eq:inversion} allows the solution to be separated into the contributions of the Marangoni poles, of the wave poles and of the continuous spectrum, shown in figure~\ref{fig:decomposition} for the same parameters. The separation identifies three temporal regimes, summarised in table~\ref{tab:regimes}. At early times, $\tau\lesssim2$, the Marangoni poles and the continuous spectrum dominate: the surfactant is redistributed by the Marangoni flow, vorticity is generated at the interface and diffuses into both fluids, and the deformable interface follows the flat-surface solution. At intermediate times the Marangoni contribution has decayed and the wave poles dominate, from $\tau\simeq5$ here: the concentration performs the weakly damped oscillation
\begin{equation}
  \theta_\Gamma\simeq2|R_w|\,\mathrm e^{-\zeta\tr}\cos(\tilde\omega\tr+\arg R_w),
  \label{eq:late-time}
\end{equation}
with $2|R_w|=0.15$. At asymptotically late times the slowest contribution dominates. The continuous spectrum decays as $\mathrm e^{-x_l\tr}\tr^{-3/2}$ (Appendix~\ref{sec:appA}), with $x_l=k^2/\Rey=3.9\times10^{-3}$, more slowly than the wave, $\zeta=0.033$, so that it eventually emerges again; for these parameters it overtakes the wave envelope only at $\tau\simeq96$, when both are below $5\times10^{-4}$. The algebraic decay $\tau^{-3/2}$ of the flat surface is the limit $x_l\to0$ of this behaviour. The deformable interface therefore does not remove the finite number of sign changes of the flat-surface relaxation, which result from the Marangoni poles together with the continuous spectrum; it adds a further, oscillatory contribution, which is the largest over the intermediate interval and whose ultimate persistence depends on the ratio of $\zeta$ to $x_l$.

\begin{table}
  \begin{center}
  \small
  \begin{tabular}{@{}p{0.26\textwidth}p{0.32\textwidth}p{0.35\textwidth}@{}}
      regime & dominant contribution & dynamics\\[3pt]
      early time, $\tau\lesssim2$ & Marangoni poles, continuous spectrum & Marangoni redistribution, vorticity generation\\
      intermediate time & wave poles & weakly damped capillary-gravity oscillation\\
      late time & slowest of $\zeta$ and $x_l$ & wave, or the continuous spectrum $\mathrm e^{-x_l\tr}\tr^{-3/2}$\\
  \end{tabular}
  \caption{Temporal regimes of the concentration for a single mode at large $\Rey$; the boundaries are given for the parameters of figure~\ref{fig:decomposition}.}
  \label{tab:regimes}
  \end{center}
\end{table}

\begin{figure}
  \centering
  \includegraphics[width=\textwidth]{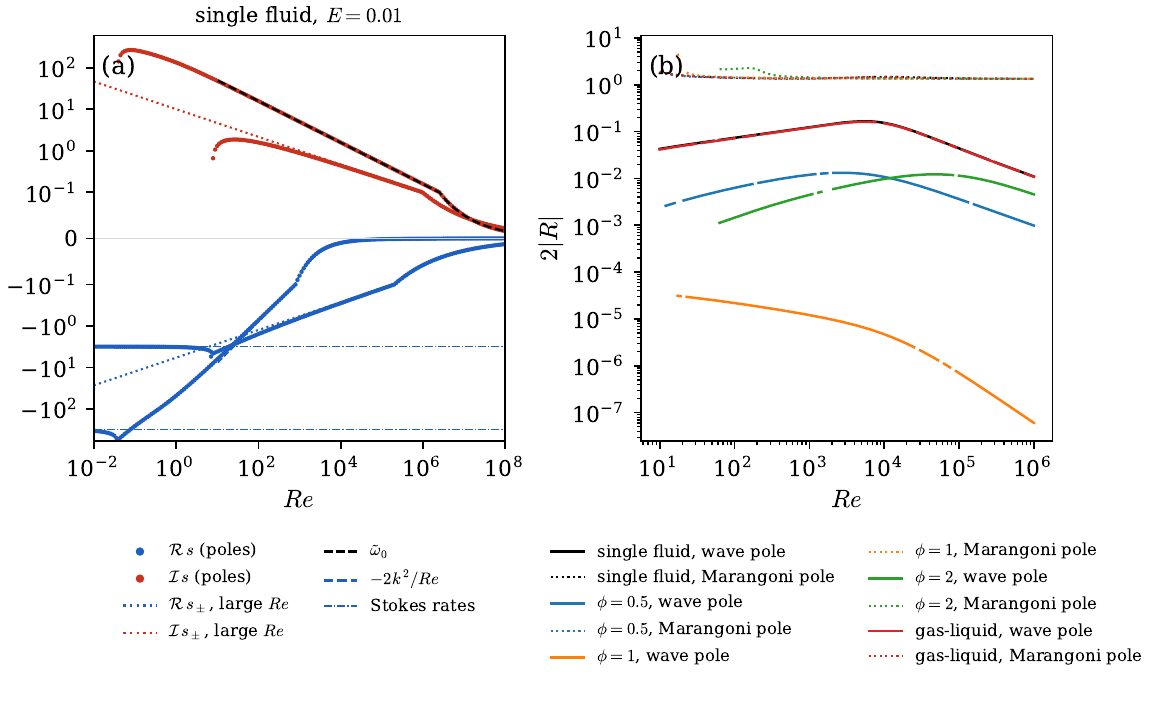}
  \caption{(a) As figure~\ref{fig:poles}(a) for $E=0.01$, for which the time scales match, \eqref{eq:crossing}, at $\Rey\simeq6\times10^6$. (b) Twice the modulus of the residue of the concentration transfer function $\Nc/(s\NN)$ at the wave pole (solid) and at the Marangoni pole (dotted), as a function of $\Rey$ for $E=0.1$: a single fluid with a free surface, two liquids with $\rho_r=0.9$ and $\phi=0.5$, $1$ and $2$, and a gas-liquid interface ($\rho_r=1.2\times10^{-3}$, $\mu_r=1.8\times10^{-2}$). The wave residue is the amplitude of the weakly damped, wave-locked oscillation of the concentration.}
  \label{fig:residues}
\end{figure}

The amplitude of the wave-locked oscillation is the wave residue, shown in figure~\ref{fig:residues}(b) as a function of $\Rey$ for several fluid pairs at $E=0.1$. For the free surface it rises with $\Rey$, peaks at $2|R_w|=0.17$ at $\Rey\simeq5\times10^3$, where $q\simeq0.97$; at this finite elasticity the maximum is displaced from the small-$E$ value $q=1/\sqrt2$ (figure~\ref{fig:residue}). It decreases beyond the maximum; the Marangoni residues remain close to the flat-surface value $2/3$ of each pole of \eqref{eq:theta-largeRe} throughout. The gas-liquid interface is indistinguishable from the free surface. For the liquid-liquid pairs the residue is an order of magnitude smaller, in proportion to $\chi^2$, and its peak moves to larger $\Rey$ with $(1+\phi)^4$. For $\phi=1$ it is smaller by four further orders of magnitude but does not vanish: the leading-order coupling $\chi$ is zero, and the residue that remains comes from the higher-order terms of $Z_c$, which vanish exactly only for $\rho_r=\mu_r=1$.

\begin{figure}
  \centering
  \includegraphics[width=\textwidth]{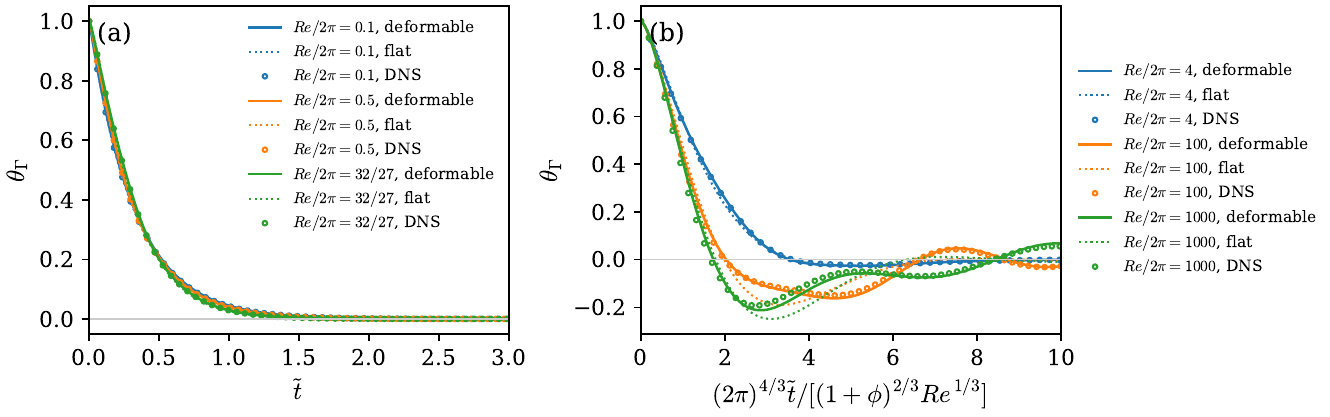}
  \caption{Decay of the surfactant distribution $J_0(kr)$ on a deformable gas-liquid interface ($E=0.1$, $\Bo=0$; solid lines, linear solution), on a flat surface (dotted lines) and in the DNS (circles; $N=256$ for $\Rey/2\pi\le4$, $N=512$ otherwise), the three Reynolds numbers of each panel being distinguished by colour: (a) $\Rey/2\pi=0.1$, $0.5$ and $32/27$ against $\tr$; (b) $\Rey/2\pi=4$, $100$ and $1000$ against the rescaled time of \eqref{eq:theta-shardt}.}
  \label{fig:decay-series}
\end{figure}

Figure~\ref{fig:decay-series} follows the decay through the regimes of figure~\ref{fig:poles}(a). Below the critical Reynolds number the decay is monotonic and the deformable and flat surfaces are indistinguishable, the exponential rate $\pi$ of the Stokes limit being approached from above as $\Rey\to0$. Above it the amplitude overshoots. At $\Rey/2\pi=4$ the single overshoot of the flat surface is reproduced; at $\Rey/2\pi=100$ the first minimum is already shallower on the deformable surface and a second oscillation follows; and at $\Rey/2\pi=1000$, which is $q=1$ for $E=0.1$, the first minimum is reduced from $-0.25$ to $-0.21$ and the subsequent oscillation has the largest amplitude. The deformable interface takes energy out of the Marangoni relaxation and stores it in the wave, from which it is returned to the surfactant distribution periodically.

\section{Direct numerical simulations}
\label{sec:num}

The linear solution of \S\S\ref{sec:analytical}-\ref{sec:single} is compared with axisymmetric direct numerical simulations of the nonlinear equations \eqref{eq:surf-full}-\eqref{eq:ic}, which also test its range of validity. The numerical method combines established ingredients and is summarised here; its details and its verification are given in the supplementary material.

\subsection{Flow solver, interface and stresses}

The two-phase incompressible Navier-Stokes equations are solved in the one-fluid formulation with the open-source library Basilisk \citep{Popinet2019basilisk}, by a second-order projection method with implicit viscosity and multigrid solution of the pressure and viscous systems \citep{popinet2009accurate,popinet2018numerical}, in axisymmetric coordinates, with $r$ the radius and the axis $r=0$ a symmetry boundary. The interface is captured with a coupled volume-of-fluid/levelset method: the volume fraction $f$, unity in the lower fluid, is advanced by geometric advection of a piecewise-linear interface construction (PLIC), and a levelset, relaxed towards the signed distance to the PLIC segments \citep{russo2000remark}, supplies the geometry required by the stress calculation. Capillary and Marangoni stresses are imposed together by the momentum-conserving integral formulation of \citet{popinet1999front,alsaud2018conservative}, in which the force on each control volume is the flux of the surface stress tensor $\sigma(\mathsf I-\mathbf n\mathbf n)\delta_\Sigma$ through its boundary, $\sigma$ being evaluated from the facet concentration by \eqref{eos}; spatial variations of $\sigma$ therefore produce the Marangoni stress within the same discrete operator as the capillary stress, and in axisymmetric coordinates the operator includes the azimuthal curvature term.

\subsection{Axisymmetric sharp surfactant transport}
\label{sec:sharp-transport}

The surfactant is stored on the reconstructed interface and transported by the sharp geometric scheme of \citet{xue2026sharp}, without adsorption, desorption or surface diffusion. In axisymmetric coordinates the PLIC segment of cell $i$ represents a conical strip, and its area and surfactant mass per radian are
\begin{equation}
  A_i=\int_{\mathcal S_i}r\,\mathrm ds=\ell_i\,\frac{r_{a,i}+r_{b,i}}{2},\qquad M_i=\Gamma_iA_i,
  \label{eq:axi-area}
\end{equation}
where $\ell_i$ is the segment length and $r_{a,i}$, $r_{b,i}$ the radii of its end points. During each directional sweep the surfactant mass crossing a cell face is the integral of a linear concentration profile, weighted by $r$, over the part of the donor segment that the sweep carries across the face,
\begin{equation}
  \Psi_{i+1/2}=\int_{\mathcal S_{i+1/2}}\big[\Gamma_{\rm up}+g_{\rm up}(s-s_{\rm up})\big]\,r(s)\,\mathrm ds ,
  \label{eq:dns-surf-flux}
\end{equation}
with $r(s)$ varying linearly along the segment, so that the radial metric enters the fluxes exactly for a straight segment. The concentration after the sweep is the updated mass divided by a corrected area,
\begin{equation}
  M_i^*=M_i^n+\Psi_{i-1/2}-\Psi_{i+1/2},\qquad
  A_i^c=A_i^n+\Phi_{i-1/2}-\Phi_{i+1/2}+\Delta t\,A_i^n\mathcal D_i,
  \label{eq:dns-area-update}
\end{equation}
where $\Phi$ is the $r$-weighted area carried across a face and $\mathcal D_i=\nabla\cdot\u-\mathbf n\cdot(\nabla\u)\cdot\mathbf n$ the surface dilation rate of the split velocity. The divergence in $\mathcal D_i$ is computed from face fluxes that carry the axisymmetric metric, so that it contains the azimuthal stretching $u_r/r$, which is absent in planar geometry. After each sweep the concentrations are multiplied by a common factor that restores the total mass $\sum_iM_i$; the global conservation of the $r$-weighted mass therefore includes this rescaling, and conservation alone is not evidence of local accuracy.

\subsection{Time step}

The time step satisfies the advective condition and the explicit capillary bound $\Delta t\le(\rho_m\Delta^3/\pi\sigma_{\max})^{1/2}$, with $\rho_m$ the mean density, and is capped at $10^{-2}$ capillary times. 

\subsection{Verification of the implementation}
\label{sec:validation-summary}

The implementation is verified in the supplementary material against five benchmarks: the expansion of a surfactant-covered circle and sphere, in planar, three-dimensional and axisymmetric geometry, which tests dilation and, in the axisymmetric case, the radial area weighting \eqref{eq:axi-area} and the azimuthal stretching; the solid-body rotation of a surfactant disturbance, which tests tangential transport; thermocapillary migration of a drop, in axisymmetric and three-dimensional geometry, which tests the variable-tension stress; and the Rayleigh-Plateau instability of a surfactant-covered thread, an axisymmetric test of the full coupling of transport, Marangoni stress and flow. In the axisymmetric expansion test the mean concentration converges to the exact dilation, with errors from $6\times10^{-3}$ to $4\times10^{-4}$ as the mesh is refined from $13$ to $100$ cells per diameter, the area-weighted error converges at first order, and the corrected mass is conserved to round-off; the error is concentrated on the two facets that touch the axis, whose area is negligible, so that the $r$-weighted diagnostics of the present simulations are unaffected but concentrations read on the axis itself are not reliable. The thread growth rates and concentration-to-displacement ratios agree with the dispersion relation of \citet{timmermans2002effect} to within $1\,\%$ and $2.5\,\%$.

\subsection{Simulation set-up and diagnostics}
\label{sec:dns-setup}

The simulations are carried out in the $(r,z)$ half-plane with the axis $r=0$ as a symmetry boundary. The Bessel mode is placed in a cylinder of radius $R$ with a free-slip, impermeable wall, $u_r=\partial u_z/\partial r=0$ at $r=R$, and $R$ is chosen at a zero of $J_1$, $kR=j_{1,3}=10.17$, so that $J_1(kR)=0$ and the mode $J_0(kr)$, its radial velocity $\propto J_1(kr)$ and its concentration gradient $\propto J_1(kr)$ satisfy the wall conditions exactly: the finite cylinder reproduces the unbounded single-mode problem, and the domain contains the central patch and two rings of the mode. The axial extent is $-R/2<z<R/2$, i.e.\ $kH=5.09$ on either side of the mean interface, with impermeable free-slip ends; the corresponding deep-water correction is of order $2\exp(-2kH)\simeq4\times10^{-5}$. A uniform $N\times N$ mesh is used with $N=256$ or $512$, that is, $158$ or $316$ cells per wavelength $\lambda=2\pi/k$. Both fluids are initially at rest, the interface is flat, and the concentration $\Gamma_0[1+\varepsilon J_0(kr)]$ is evaluated at the PLIC segment centroids, with $\varepsilon=0.1$; this is the single mode \eqref{eq:single-ic}. The solver works in the capillary units $k=\rho\LO=\Gamma_0=1$, $\sigma_0=1/2$; the dimensionless groups of \S\ref{sec:nondim} are imposed through the viscosity $\mu\LO=\Oh(\rho\LO\sigma_0/k)^{1/2}$ with $\Oh=(2\pi E/\Rey)^{1/2}$, the elasticity $\beta=E\sigma_0$ and the ratios $\rho_r$, $\mu_r$; gravity is zero ($\Bo=0$), as in the figures of \S\ref{sec:single}. The time step is capped at $10^{-2}$ capillary times and is otherwise limited as described in \S\ref{sec:num}; the solver tolerance parameter is $10^{-8}$.

Each case is computed twice, with the mean interface placed $0.4\Delta$ and $0.6\Delta$ above an axial grid line, and the two results are averaged. Exact alignment of the interface with a grid line or a row of cell centres is thereby avoided, and the mean reduces the sensitivity of the results to the position of the interface within a cell; the two members of each pair are also reported separately where they differ visibly. The interface displacement and the concentration perturbation are measured as projections on the mode,
\begin{equation}
\begin{gathered}
  a_{\rm DNS}(t)=\frac{\sum_i f_i(t)J_0(kr_i)\,r_i\Delta^2}{\sum_i J_0^2(kr_i)\,r_i\Delta^2/2H},\\
  \gamma_{\rm DNS}(t)=
  \frac{\sum_i[\Gamma_i(t)-\Gamma_0]J_0(kr_{\Sigma,i})W_i}{\sum_i J_0^2(kr_{\Sigma,i})W_i},
  \qquad W_i=\ell_i|n_{z,i}|\,r_{\Sigma,i}=A_i|n_{z,i}|,
\end{gathered}
  \label{eq:dns-amplitude}
\end{equation}
where the first sum runs over all cells and uses $\int_0^RJ_0(kr)\,r\,\mathrm dr=0$ at a zero of $J_1$, so that the mean position of the interface drops out, $r_{\Sigma,i}$ is the segment-centroid radius and $W_i$ the radial extent of segment $i$ weighted once by its radius, $A_i$ being the area per radian \eqref{eq:axi-area}; the small static value of $a_{\rm DNS}(0)$ left by the discrete initialisation is subtracted. In the variables of \S\ref{sec:single}, $\theta_\Gamma=\gamma_{\rm DNS}/\varepsilon\Gamma_0$ and $\theta_\eta=ka_{\rm DNS}/2\pi\varepsilon$. The discrete initialisation changes the mode amplitude slightly from its nominal value, so the linear solution is evaluated with the measured $\gamma_{\rm DNS}(0)$; no later DNS data are fitted. The reference is the numerical inversion of \eqref{eq:Gamma-sol}-\eqref{eq:eta-sol}, which retains the branch-cut contribution. We also monitor the total surfactant mass, the lower-fluid volume, the concentration extrema and the kinetic energy.

\begin{table}
  \begin{center}
  \footnotesize
  \setlength{\tabcolsep}{2.6pt}
  \begin{tabular}{lllcccccl}
      set & fluids & $\rho_r$, $\mu_r$ & $\Rey/2\pi$ & $E$ & $\Bo$ & $\varepsilon$ & $N$ & figure\\[3pt]
      Stokes & GL; $\phi=1$ & GL; 0.9, 1.1 & 0.1 & 0.1 & 0 & 0.1 & 256 & \ref{fig:dns-stokes}\\
      regimes & gas-liquid & GL & 0.1, 0.5, $32/27$, 4 & 0.1 & 0 & 0.1 & 256 & \ref{fig:decay-series}, \ref{fig:dns-regimes}\\
      regimes & gas-liquid & GL & 100, $10^3$, $10^4/2\pi$ & 0.1 & 0 & 0.1 & 256, 512 & \ref{fig:decay}, \ref{fig:decay-series}, \ref{fig:dns-regimes}\\
      resonance & gas-liquid & GL & 190 & 0.3 & 0 & 0.1 & 256, 512 & \ref{fig:dns-resonance}\\
      resonance & $\phi=0.5$ & 0.9, 0.28 & 190 & 0.3 & 0 & 0.1 & 256, 512 & \ref{fig:dns-resonance}\\
      resonance & $\phi=1$, $2$ & 0.9; 1.1, 4.4 & 190 & 0.3 & 0 & 0.1 & 256, 512 & \ref{fig:dns-resonance}\\
      Bond & gas-liquid & GL & 100 & 0.1 & $4\pi^2$ & 0.1 & 256 & \ref{fig:dns-bond}\\
      localized & gas-liquid & GL & $\Oh=0.04$, $ka=2$ & 0, 0.28 & 0 & 0.1; $a_0/a=0.03$ & 256, 512 & \ref{fig:config-cp}, \ref{fig:deposit}\\
  \end{tabular}
  \caption{Simulated cases; every case is run at the two grid offsets. GL denotes the gas-liquid pair $\rho_r=1.2\times10^{-3}$, $\mu_r=1.8\times10^{-2}$ ($\phi=4.6\times10^{-3}$). The Ohnesorge number $\Oh=(2\pi E/\Rey)^{1/2}$ sets the viscosity in the solver: $\Oh=0.04$ in the resonance and localized sets, $0.032$ at $\Rey/2\pi=100$, $0.01$ at $\Rey/2\pi=1000$ and $8\times10^{-3}$ at $\Rey=10^4$. The localized set is described in \S\ref{sec:deposit}.}
  \label{tab:dns-cases}
  \end{center}
\end{table}

Table~\ref{tab:dns-cases} lists the cases. They test the statements of \S\ref{sec:single} that go beyond the flat-surface problem: that the interface does not move in the Stokes limit, that the decay changes from monotonic to oscillatory at $\Rey/2\pi\simeq32/27$ and departs from the flat-surface solution as $\Rey$ grows, and that near the time-scale matching \eqref{eq:crossing} the surfactant initiates a wave, of polarity set by $\phi$, which carries the concentration. The thinnest layer to be resolved is the oscillatory boundary layer of the wave, of thickness $(\mu\LO/\rho\LO\omega_0)^{1/2}=\lambda(\Rey\,\tilde\omega_0)^{-1/2}$, which is $0.03\lambda$ at $\Oh=0.04$, or 4-8 cells for $N=256$-$512$. The resonance set is therefore run at the moderate $\Rey=1.2\times10^3$ and placed near $q=1$ by the choice $E=0.3$; the cases at larger $\Rey$ ($\Oh\le0.01$), used only to compare the evolution of the concentration with the linear solution in figures~\ref{fig:decay} and \ref{fig:decay-series}, are less well resolved, and the agreement there should be read with this in mind.

\section{Comparison with direct numerical simulations}
\label{sec:dns-results}

\subsection{Stokes limit}

\begin{figure}
  \centering
  \includegraphics[width=\textwidth]{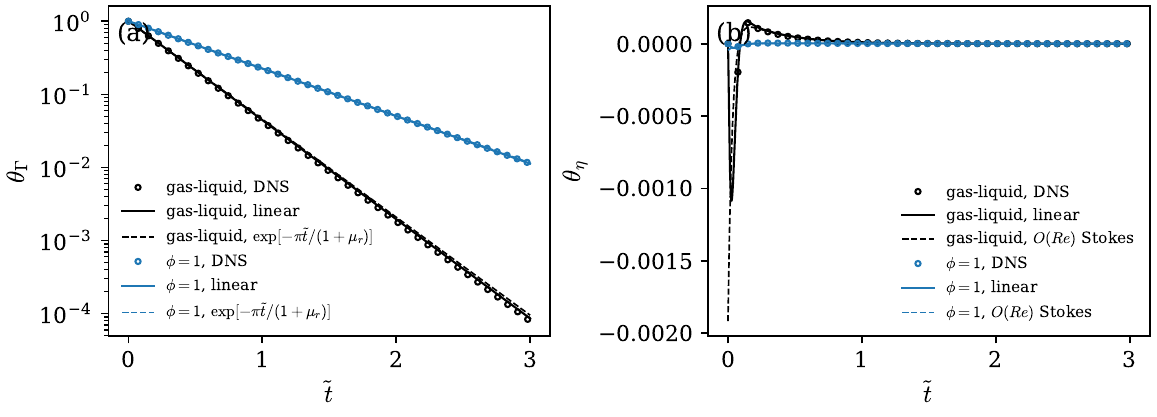}
  \caption{Stokes limit, $\Rey/2\pi=0.1$, $E=0.1$, for a gas-liquid interface and for two liquids with $\phi=1$. (a) Concentration amplitude $\theta_\Gamma$: DNS (symbols, mean of the two grid offsets), linear solution (solid) and the Stokes-limit exponential $\exp[-\pi\tr/(1+\mu_r)]$ (dashed). (b) Interface displacement $\theta_\eta$: DNS (symbols), linear solution (solid) and the first-order-in-$\Rey$ prediction \eqref{eq:stokes-ORe} (dashed).}
  \label{fig:dns-stokes}
\end{figure}

Figure~\ref{fig:dns-stokes} tests the two statements of \S\ref{sec:stokes}. The concentration decays exponentially, at the rate $\pi/(1+\mu_r)$ of \eqref{eq:stokes-rates}, which is $3.09$ for the gas-liquid interface and $1.49$ for $\phi=1$; the rates fitted to the DNS over $0.3<\tr<2$ are $3.16$ and $1.49$, and the DNS follow the linear solution to within $4\times10^{-3}$ and $9\times10^{-4}$ in $\theta_\Gamma$ respectively. The small excess over the pure exponential is the inertial shift of the real pole, $(1+\rho_r)\Rey/8k(1+\mu_r)^2$ in relative terms, which the linear solution contains. The interface moves by $10^{-3}$ in units of $\varepsilon\lambda$ for the gas-liquid interface, the DNS maximum of $|\theta_\eta|$ being $1.06\times10^{-3}$ against $1.08\times10^{-3}$ in the linear solution and in agreement with the first-order prediction \eqref{eq:stokes-ORe}, which is $O(\Rey)$ and directed downward on the axis, beneath the concentration maximum; for $\phi=1$ the displacement is $3\times10^{-5}$ in both. The Marangoni flow at small Reynolds number therefore leaves the interface flat to the accuracy of the simulations, and the residual displacement is the inertial correction, proportional to $\Rey(1-\rho_r)$.

\subsection{From monotonic to oscillatory decay}

\begin{figure}
  \centering
  \includegraphics[width=\textwidth]{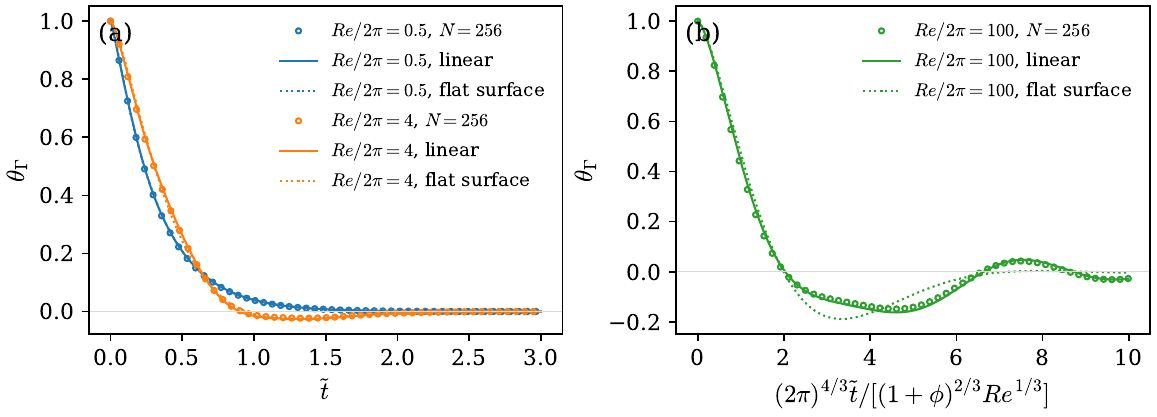}
  \caption{Decay of the concentration on a gas-liquid interface at $E=0.1$: DNS (symbols; circles $N=256$, squares $N=512$), linear solution for the deformable interface (solid) and for the flat surface (dotted). (a) $\Rey/2\pi=0.5$ and $4$ against $\tr$; (b) $\Rey/2\pi=100$ against the rescaled time of \eqref{eq:theta-shardt}.}
  \label{fig:dns-regimes}
\end{figure}

Figure~\ref{fig:dns-regimes} follows the decay through the regimes of figure~\ref{fig:poles}(a). Below the critical Reynolds number ($\Rey/2\pi=0.5$) the decay is monotonic and the deformable and flat surfaces coincide; at $\Rey/2\pi=4$ the amplitude overshoots once, as in figure~6 of \citet{Shardt2016oscillatory}, and at $\Rey/2\pi=100$ it overshoots and the deformable interface departs from the flat surface, the first minimum being shallower and followed by a further oscillation. The DNS reproduce the linear solution to within $3\times10^{-3}$, $6\times10^{-3}$ and $2\times10^{-2}$ in $\theta_\Gamma$ at the three Reynolds numbers for $N=256$; the largest difference occurs at the first minimum of the $\Rey/2\pi=100$ case, $-0.147$ against $-0.162$, where the boundary layer is thinnest, and it decreases to $-0.153$, a maximum difference of $1.3\times10^{-2}$, on refinement to $N=512$.

\subsection{Initiated wave, polarity and wave-locked oscillation}

\begin{figure}
  \centering
  \includegraphics[width=\textwidth]{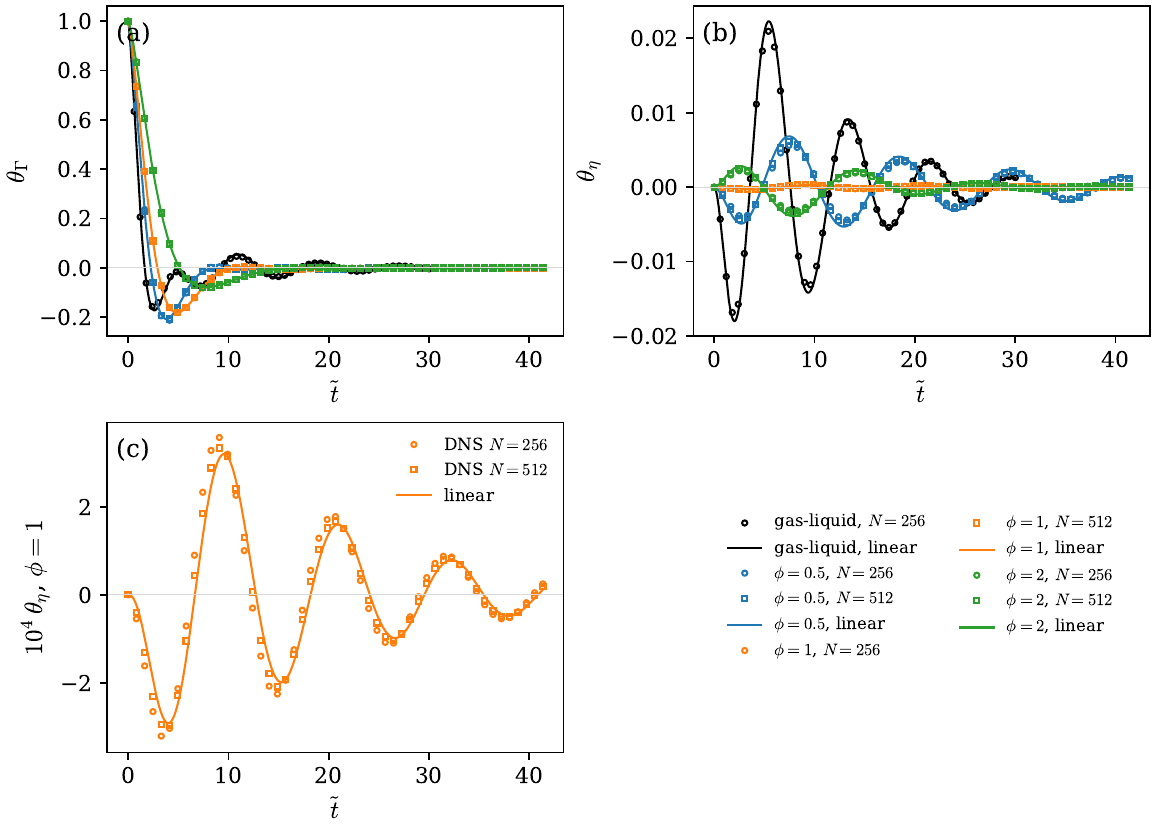}
  \caption{The resonance set, $\Rey=1.2\times10^3$, $E=0.3$: a gas-liquid interface (black) and two liquids with $\phi=0.5$ (blue), $1$ (orange) and $2$ (green). (a) Concentration amplitude and (b) interface displacement: DNS (circles $N=256$, squares $N=512$) and linear solution (solid lines). (c) The displacement for $\phi=1$, magnified by $10^4$.}
  \label{fig:dns-resonance}
\end{figure}

Figure~\ref{fig:dns-resonance} shows the cases nearest the time-scale matching \eqref{eq:crossing}. On the gas-liquid interface the surfactant initiates a wave: the interface on the axis, beneath the concentration maximum, first moves downward, to $\theta_\eta=-0.017$ ($-0.018$ in the linear solution), and then swings to $+0.021$ ($+0.022$), in units of $\varepsilon\lambda$, and the concentration, after its first minimum, is carried by this wave through the four periods simulated instead of decaying monotonically. For the two liquids the displacement is smaller by a factor of three to six, as the residues of figure~\ref{fig:residues}(b) predict, and its polarity reverses between $\phi=0.5$ and $\phi=2$: the first extremum of $\theta_\eta$ is $-4.3\times10^{-3}$ ($-4.9\times10^{-3}$ in the linear solution) for $\phi=0.5$ and $+2.5\times10^{-3}$ ($+2.8\times10^{-3}$) for $\phi=2$, while for $\phi=1$ it is $-3\times10^{-4}$ in both, an order of magnitude smaller (figure~\ref{fig:dns-resonance}c). The displacement at $\phi=1$ is small but not zero: these liquids have $\rho_r=0.9$, so the coupling $Z_c$ vanishes only at leading order in $\Rey^{-1/2}$, and the DNS reproduce the residual displacement of the linear solution. The sign of $\chi$ therefore sets the direction in which a surfactant disturbance moves the interface. The concentration amplitudes agree with the linear solution to within $8\times10^{-3}$ for the three liquid pairs at $N=512$ and $2.5\times10^{-2}$ for the gas-liquid interface at $N=256$, and the $(1+\phi)^{-2/3}$ slowing of the relaxation for $\phi=2$ is visible in figure~\ref{fig:dns-resonance}(a).

\subsection{Finite Bond number}
\label{sec:bond}

\begin{figure}
  \centering
  \includegraphics[width=\textwidth]{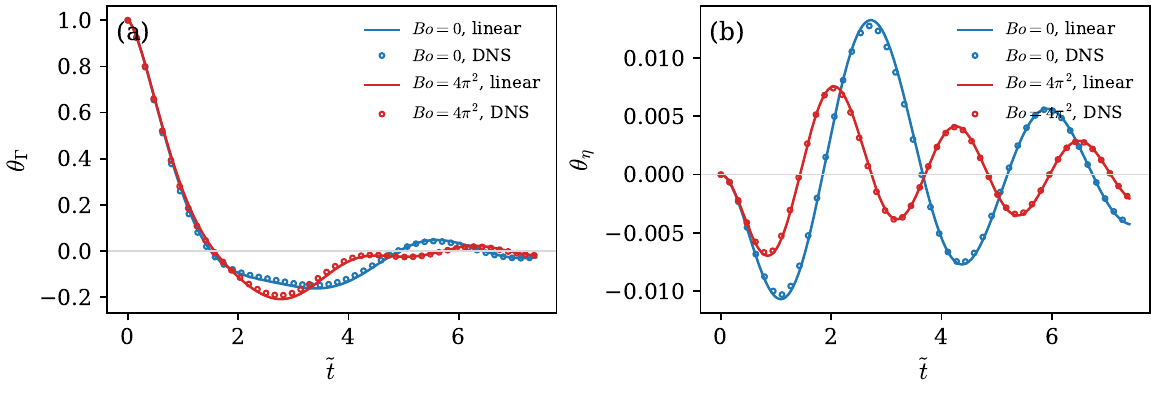}
  \caption{Effect of gravity on a gas-liquid interface at $\Rey/2\pi=100$, $E=0.1$: $\Bo=0$ (blue) and $\Bo=4\pi^2$, for which gravity and capillarity contribute equally to the stiffness of the mode (red). (a) Concentration amplitude and (b) interface displacement: linear solution (lines) and DNS at $N=256$ (circles).}
  \label{fig:dns-bond}
\end{figure}

Gravity enters the problem only through the stiffness $G=(\Bo+k^2)/E$. Figure~\ref{fig:dns-bond} compares $\Bo=0$ with $\Bo=4\pi^2$, for which gravity doubles the stiffness of the mode $k=2\pi$ and raises the wave frequency by $\sqrt2$. The initiated displacement is reduced, from a largest value of $1.3\times10^{-2}$ to $7.6\times10^{-3}$ in units of $\varepsilon\lambda$. Gravity acts in two ways: the prefactor $k^2/G$ of \eqref{eq:residue} and \eqref{eq:eta-largeRe} halves, and the parameter $q$, proportional to $(1+\Bo/4\pi^2)^{-3/4}$, decreases, which moves the case along the resonance factor of \eqref{eq:residue}; and the wave-locked oscillation of the concentration is correspondingly weaker and faster; the early relaxation of the concentration, which is controlled by the Marangoni poles, is unchanged. The DNS follow the linear solution in both cases, with largest differences of $2\times10^{-2}$ in $\theta_\Gamma$ and $3\times10^{-4}$ in $\theta_\eta$ at $\Bo=4\pi^2$.

\section{Localized disturbances}
\label{sec:deposit}

\begin{figure}
  \centering
  \includegraphics[width=\textwidth]{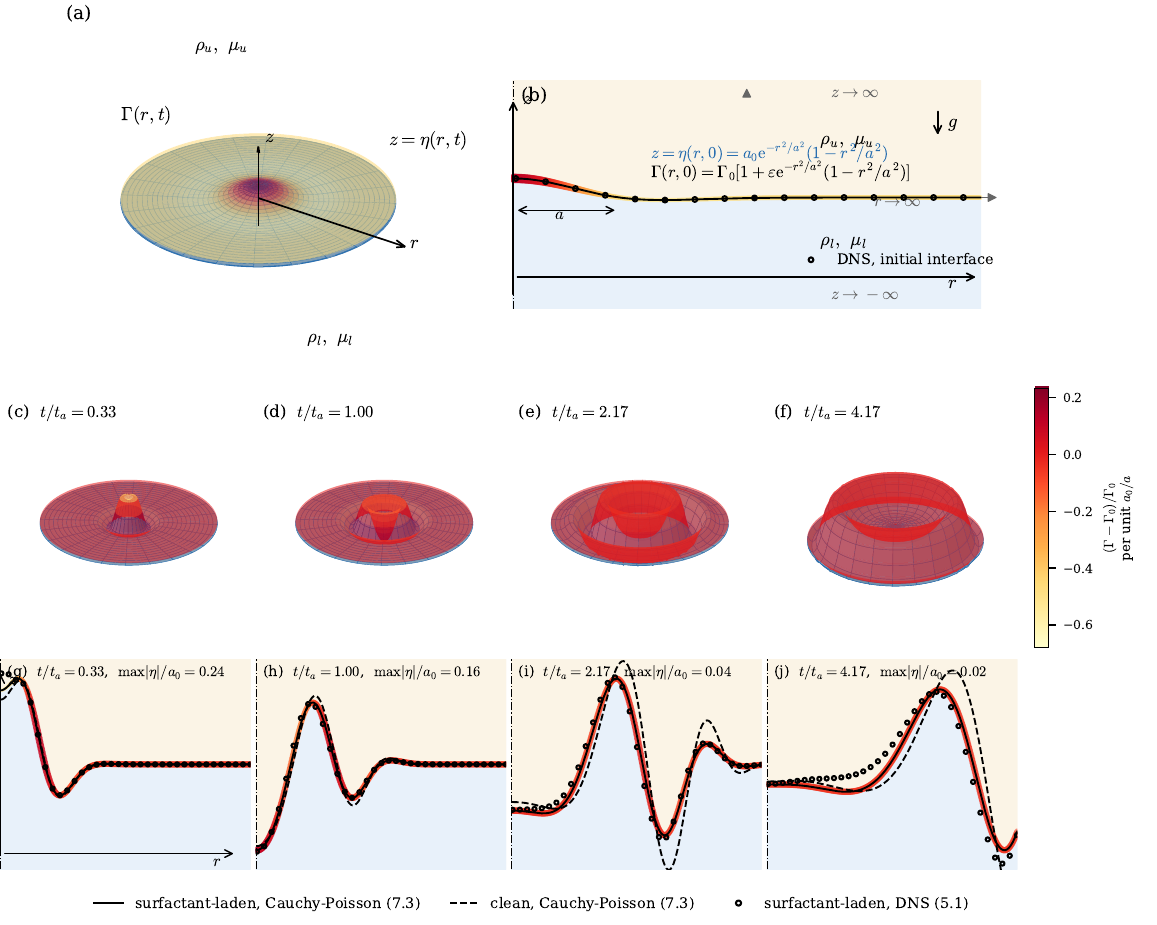}
  \caption{The zero-mass redistribution, drawn as in figure~\ref{fig:config}: (a) the interface $z=\eta(r,t)$ with the surfactant layer (colour and thickness proportional to $\Gamma$); (b) the axisymmetric section, with the initial displacement $\eta(r,0)=a_0\mathrm e^{-r^2/a^2}(1-r^2/a^2)$ and concentration $\Gamma(r,0)=\Gamma_0[1+\varepsilon\mathrm e^{-r^2/a^2}(1-r^2/a^2)]$ of width $a$. The open circles in (b) are the interface of the DNS at $t=0$. (c-f) Three-dimensional views and (g-j) axisymmetric sections of the surfactant-laden interface released from the displacement \eqref{eq:deposit-ic} ($\varepsilon=0$, parameters of figure~\ref{fig:deposit}) at $t/t_a=0.33$, $1.0$, $2.2$ and $4.2$, for $0<r<8a$: Cauchy-Poisson solution \eqref{eq:deposit-solution} for the surfactant-laden interface (solid line with the surfactant layer) and for the clean interface (black dashed line), and the surfactant-laden interface of the DNS at $N=256$ (open circles). Each snapshot is scaled to the maximum displacement of the surfactant-laden solution, quoted in the section panel as $\max|\eta|/a_0$; the colour of the surface layer and the colour bar give the surfactant perturbation $(\Gamma-\Gamma_0)/\Gamma_0$ per unit $a_0/a$ drawn along by the wave.}
  \label{fig:config-cp}
\end{figure}

The Cauchy-Poisson problem proper is the response to a localized disturbance, which excites every mode of \S\ref{sec:single} at once; the Hankel integral \eqref{eq:hankel} of the single-mode solution, which retains the continuous spectrum of each mode, gives it directly. We consider the initial distribution of width $a$
\begin{equation}
  \eta(r,0)=a_0\,\mathrm e^{-r^2/a^2}\Big(1-\frac{r^2}{a^2}\Big),\qquad
  \Gamma(r,0)=\Gamma_0\Big[1+\varepsilon\,\mathrm e^{-r^2/a^2}\Big(1-\frac{r^2}{a^2}\Big)\Big],
  \label{eq:deposit-ic}
\end{equation}
whose profile has zero radial integral: it displaces no volume and adds no net surfactant, and is a localized redistribution, a central patch surrounded by a ring of opposite sign. Its Hankel transform is
\begin{equation}
  \mathcal H_0\Big\{\mathrm e^{-r^2/a^2}\Big(1-\frac{r^2}{a^2}\Big)\Big\}=\frac{a^4k^2}{8}\,\mathrm e^{-a^2k^2/4},
  \label{eq:deposit-spectrum}
\end{equation}
which vanishes at $k=0$ and has its maximum at $k=2/a$; the weight with which the modes enter the integral \eqref{eq:hankel} is $k\hat f(k)\propto k^3\mathrm e^{-a^2k^2/4}$, which peaks at $k=\sqrt6/a$, so that the modes $2/a\lesssim k\lesssim\sqrt6/a$ dominate. To linear order the total surfactant on the interface is conserved, the change of interfacial area produced by the deformation being of second order in the amplitude; the redistribution therefore keeps its zero net excess. The solution is
\begin{equation}
  \eta(r,t)=\int_0^\infty\big[\eh_0(k)\,\theta_\eta^{\eta}(k,t)+\Gh_0(k)\,\theta_\eta^{\Gamma}(k,t)\big]J_0(kr)\,k\,\mathrm dk ,
  \label{eq:deposit-solution}
\end{equation}
and likewise for $\Gamma$, where $\theta^{\eta}_\eta(k,t)$ and $\theta^{\Gamma}_\eta(k,t)$ are the inverse Laplace transforms of the two transfer functions of \eqref{eq:eta-sol} for the mode $k$. Because the scaling of \S\ref{sec:nondim} is built on the wavelength, each mode has its own groups: with the dimensional fluid properties fixed, the mode $k$ has $\lambda_k=2\pi/k$, $\Rey_k=\rho\LO\beta\lambda_k/\mu\LO^2$, $t^*_k=\lambda_k\mu\LO/\beta$ and $\Oh_k=\mu\LO(k/\rho\LO\sigma_0)^{1/2}$, while $E$, $\rho_r$ and $\mu_r$ are common to all modes; the transfer functions are evaluated with these groups at the Laplace variable $s\,t^*_k$, and the amplitudes are rescaled from units of $\varepsilon\lambda_k$ to units of $a$. The integral is evaluated on a grid refined towards $k=0$, with each Laplace transform inverted numerically \citep{Cohen2007}.

All cases are for a gas-liquid interface with $\Bo=0$, with the viscosity chosen so that the mode $k=2/a$ has $\Oh=0.04$, the resolution limit of the simulations. We compute a clean interface released from the displacement with $a_0=0.03a$ and $\varepsilon=0$; the same displacement on a surfactant-laden interface with $E=0.28$, the elasticity of largest damping, $q=\sqrt2$, for the mode $k=2/a$; and the surfactant redistribution $\varepsilon=0.1$ on a flat interface ($a_0=0$) with the same $E$. The simulations use the axisymmetric set-up of \S\ref{sec:dns-setup} in a cylinder of radius $12a$ and height $12a$, which the waves do not reach within the simulated time $6.7t_a$, where $t_a=(\rho\LO a^3/\sigma_0)^{1/2}$ is the capillary time of the disturbance, with $21$ and $43$ cells per $a$ at $N=256$ and $512$.

\begin{figure}
  \centering
  \includegraphics[width=\textwidth]{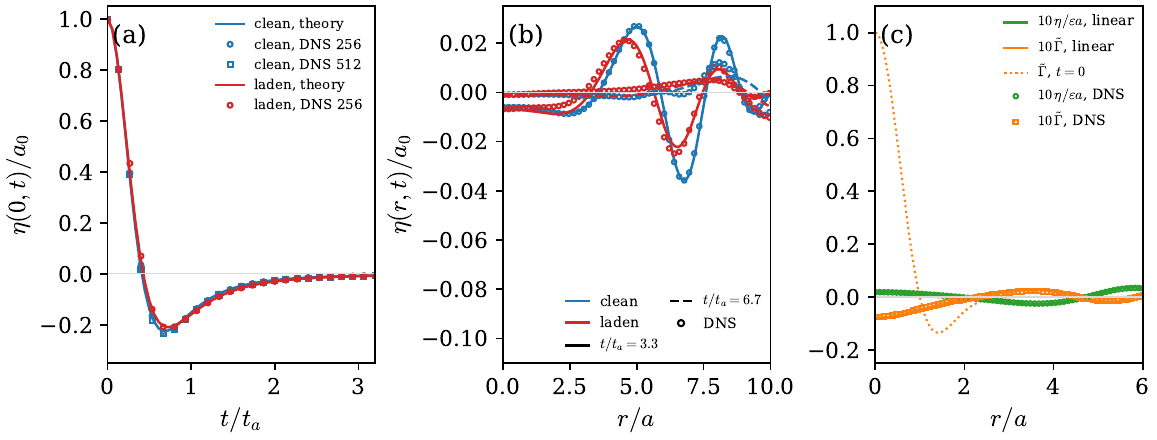}
  \caption{The zero-mass disturbance \eqref{eq:deposit-ic}, gas-liquid interface, $\Oh=0.04$ and $E=0.28$ at $k=2/a$. (a) Displacement of the interface on the axis for the clean (blue) and surfactant-laden (red) interface released from \eqref{eq:deposit-ic} with $a_0=0.03a$: linear Hankel-integral solution \eqref{eq:deposit-solution} (lines) and DNS (circles $N=256$; squares $N=512$, clean interface). (b) Interface profiles at $t/t_a=3.3$ (solid) and $6.7$ (dashed) for the same two cases, DNS at $N=256$ as circles. (c) Surfactant redistribution on a flat interface, $\varepsilon=0.1$: profiles of the initiated displacement $\eta/\varepsilon a$ (green) and of the concentration $\tilde\Gamma$ (orange) at $t/t_a=3.3$, both magnified ten times, with the initial concentration dotted; DNS at $N=512$ as symbols.}
  \label{fig:deposit}
\end{figure}

Figures~\ref{fig:config-cp} and \ref{fig:deposit} show the response to the displacement \eqref{eq:deposit-ic}. The centre of the disturbance falls, overshoots below the undisturbed level at $t\simeq0.7t_a$ and relaxes, while the disturbance is radiated outward as an annular packet of capillary waves (figure~\ref{fig:config-cp}c-j), which widens as it travels, because the modes it contains are dispersive, and whose crest is followed by a train of shorter ripples. The concentration is drawn along by the wave: where the interface is stretched, as at the centre while it first falls, the surfactant is depleted, and where it is compressed the surfactant accumulates, so that the surfactant layer of figure~\ref{fig:config-cp} carries the imprint of the packet. The surfactant leaves the first motion of the centre almost unchanged, because the Marangoni damping needs about a period of the dominant mode to act (figure~\ref{fig:deposit}a). It then acts on every mode of the packet, as the resonant damping does on a single mode: in the sections of figure~\ref{fig:config-cp}(g-j) and the profiles of figure~\ref{fig:deposit}(b) the surfactant-laden packet falls progressively below the clean one, shown dashed, and the ripples that trail the clean packet are much weaker on the surfactant-laden interface. A surfactant thus damps the travelling packet of the Cauchy-Poisson problem as it damps a standing wave.

The surfactant redistribution \eqref{eq:deposit-ic} on an initially flat interface (figure~\ref{fig:deposit}c) relaxes within about $t_a$: its central patch and the ring of opposite sign spread into each other, the interface is drawn down on the axis, and the energy that remains is carried away as a weak annular wave, along which the concentration oscillates with small amplitude. A clean interface would not move at all.

The simulations reproduce these features (figures~\ref{fig:config-cp} and \ref{fig:deposit}): the fall and overshoot of the centre, the shape and position of the radiated packet and its reduction by the surfactant, and, for the surfactant redistribution, the initiated displacement and the spreading of the concentration. The analytical solution \eqref{eq:deposit-solution} of the Cauchy-Poisson problem is thereby verified by direct numerical simulation. The surfactant-laden case at $N=512$ gives an overshoot of $-0.207a_0$ on the axis, against $-0.210a_0$ in the linear solution, and halves the largest difference of the axis displacement from the linear solution, from $0.077a_0$ at $N=256$ to $0.038a_0$ over the interval $t\le1.9t_a$ that this simulation covers.

\section{Concluding remarks}
\label{sec:conclusions}

We have solved the linear Cauchy-Poisson problem for an axisymmetric surfactant-laden interface between two viscous fluids, with independently prescribed distributions of surfactant and of interface displacement, in closed form in the Hankel-Laplace domain, and verified the solution by direct numerical simulation. The resonant interaction of the longitudinal and transverse branches, and the damping it produces, are known from normal-mode analyses; the new content is the transient coupling between surfactant redistribution and interface deformation, which answers the question with which we began.

How does a non-uniform surfactant distribution generate waves? Through a single coupling, the off-diagonal impedance $Z_c$ of \eqref{eq:impedance}, by which a tangential surface motion exerts a normal stress on the interface. In Stokes flow it vanishes: the pressure and the viscous normal stress of the Marangoni flow cancel exactly on the interface, for any viscosity ratio, the surfactant relaxes at the flat-surface rate divided by $1+\mu_r$, and the interface does not move. The coupling is therefore inertial; at small $\Rey$ it is proportional to $\Rey(1-\rho_r)$, and at large $\Rey$, at leading order for thin boundary layers, to the difference $1-\phi$ of the shear stresses of the two vortical boundary layers, so that the polarity of the initiated wave reverses when $\rho\UP\mu\UP$ exceeds $\rho\LO\mu\LO$. It vanishes identically only for fluids of equal density and equal viscosity. The amplitude of the wave-locked oscillation is $(E\chi^2/(1+\Bo/k^2))/(1-\sqrt2q+q^2)$ at small viscosity: varied through the Reynolds number at fixed elasticity it is largest at $q=1/\sqrt2$, where it equals $2E\chi^2/(1+\Bo/k^2)$, and varied through the elasticity at fixed Ohnesorge number it is largest at $q=1$; the familiar maximum of the damping lies at $q=\sqrt2$, although excitation and damping share the resonance factor $(1-\sqrt2q+q^2)^{-1}$.

How do the waves then govern the surfactant? The concentration is carried by the wave it has initiated. After an early regime of Marangoni redistribution, in which the deformable interface follows the flat-surface solution of \citet{Shardt2016oscillatory}, the wave poles dominate an intermediate regime in which the concentration performs weakly damped oscillations at the wave frequency; only at much later times does the continuous spectrum, decaying as $\mathrm e^{-x_l\tr}\tr^{-3/2}$, emerge again. For a localized displacement the surfactant damps the radiated packet, as it damps a single mode, and a localized surfactant redistribution initiates a wave on an interface that a clean interface would leave flat.

These results are linear, axisymmetric and for an insoluble surfactant without surface diffusion. Finite amplitudes and surfactant deposits of finite mass, whose spreading on a clean interface is the nonlinear problem of \citet{Jensen1995spreading}, soluble surfactants and finite depth are natural extensions; the impedance formulation \eqref{eq:impedance} extends directly to the last.

\appendix
\section{Inversion of the Laplace transform}
\label{sec:appA}
\usetikzlibrary{arrows.meta,decorations.markings,decorations.pathreplacing,calc}
\definecolor{k1}{HTML}{2A78D6}  % blue
\definecolor{k2}{HTML}{EB6834}  % orange
\definecolor{k3}{HTML}{1BAF7A}  % aqua
\definecolor{k4}{HTML}{EDA100}  % yellow
\definecolor{k5}{HTML}{E87BA4}  % magenta
\definecolor{k6}{HTML}{008300}  % green
\definecolor{k7}{HTML}{4A3AA7}  % violet
\definecolor{k8}{HTML}{C1272D}  % red
\begin{figure}[h!]
\centering
\begin{tikzpicture}[scale=0.75, transform shape, >=Stealth, line cap=round, cont/.style={very thick, draw=#1, postaction={decorate}, decoration={markings, mark=at position 0.5 with {\arrow{Stealth[length=2.8mm]}}}}, seg/.style={circle, draw=#1, text=#1, inner sep=1pt, font=\small\bfseries, fill=white}, pole/.style={circle, draw, inner sep=1.6pt, fill=white}, bp/.style={circle, draw, fill, inner sep=1.4pt}]
  \def\R{5.2}      % radius of the large arcs
  \def\c{1.0}      % Bromwich abscissa
  \def\xu{2.9}     % x_u = -s_{b,u}
  \def\xl{1.0}     % x_l = -s_{b,l}
  \def\gap{0.12}   % half-width of the cut (drawn exaggerated)
  \def\dl{0.25}    % radius of the small circle
  \pgfmathsetmacro{\aone}{acos(\c/\R)}
  \pgfmathsetmacro{\ym}{\R*sin(\aone)}
  \pgfmathsetmacro{\ag}{asin(\gap/\R)}
  \pgfmathsetmacro{\ac}{180-asin(\gap/\dl)}
  \pgfmathsetmacro{\bx}{\R*cos(\ag)}
  % axes
  \draw[->] (-\R-0.4,0) -- (2.7,0) node[right] {$\mathcal R\,s$};
  \draw[->] (0,-\ym-0.6) -- (0,\ym+0.6) node[above] {$\mathcal I\,s$};
  % contour
  \draw[cont=k1] (\c,-\ym) -- (\c,\ym);
  \draw[cont=k2] (\c,\ym) arc[start angle=\aone, end angle=180-\ag, radius=\R];
  \draw[cont=k3] (-\bx,\gap) -- (-\xu,\gap);
  \draw[cont=k4] (-\xu,\gap) -- ({-\xl+\dl*cos(\ac)},\gap);
  \draw[very thick, draw=k5, postaction={decorate},
        decoration={markings, mark=at position 0.5 with {\arrow{Stealth[length=2.2mm]}}}]
        ({-\xl+\dl*cos(\ac)},\gap) arc[start angle=\ac, end angle=-\ac, radius=\dl];
  \draw[cont=k6] ({-\xl+\dl*cos(\ac)},-\gap) -- (-\xu,-\gap);
  \draw[cont=k7] (-\xu,-\gap) -- (-\bx,-\gap);
  \draw[cont=k8] (-\bx,-\gap) arc[start angle=180+\ag, end angle=360-\aone, radius=\R];
  % segment numbers
  \node[seg=k1] at (\c+0.45,1.6) {1};
  \node[seg=k2] at ({\R*cos(125)-0.25},{\R*sin(125)+0.25}) {2};
  \node[seg=k3] at (-4.0,0.55) {3};
  \node[seg=k4] at (-1.95,0.55) {4};
  \node[seg=k5] at (-\xl+0.75,0.42) {5};
  \node[seg=k6] at (-1.95,-0.55) {6};
  \node[seg=k7] at (-4.0,-0.55) {7};
  \node[seg=k8] at ({\R*cos(235)-0.25},{\R*sin(235)-0.25}) {8};
  \draw (\c,0.08) -- (\c,-0.08) node[below right] {$c$};
  \node[right] at (0.12,-0.25) {$0$};
  % branch points
  \node[bp] at (-\xu,0) {};
  \node[bp] at (-\xl,0) {};
  \node[below=4pt] at (-\xu,-\gap) {$s_{b,u}$};
  \node[below=4pt] at (-\xl,-\gap) {$s_{b,l}$};
  \draw[decorate, decoration={brace, amplitude=4pt, mirror}] (-\bx,-1.1) -- (-\xu-0.05,-1.1)
        node[midway, below=5pt] {$x>x_u$};
  \draw[decorate, decoration={brace, amplitude=4pt, mirror}] (-\xu+0.05,-1.1) -- (-\xl+0.2,-1.1)
        node[midway, below=5pt] {$x_l<x<x_u$};
  % poles: wave pair near the imaginary axis, Marangoni pair further left
  \node[pole, label={left:$-\zeta+\mathrm i\tilde\omega$}] at (-0.45,3.4) {};
  \node[pole, label={left:$-\zeta-\mathrm i\tilde\omega$}] at (-0.45,-3.4) {};
  \node[pole, label={left:$s_+$}] at (-2.4,2.6) {};
  \node[pole, label={left:$s_-$}] at (-2.4,-2.6) {};
\end{tikzpicture}
\caption{Contour used to invert the Laplace transforms \eqref{eq:Gamma-sol}-\eqref{eq:eta-sol}, drawn for $x_l<x_u$ and closed in the left half of the plane of the Laplace variable $s$. The branch points are $s_{b,l}=-x_l$ and $s_{b,u}=-x_u$ of \eqref{eq:branchpoints}, with $x_l=k^2/\Rey$ and $x_u=k^2\mu_r/(\Rey\,\rho_r)$. The poles on the physical sheet are shown for the oscillatory regime of \S\ref{sec:poles}: the capillary-gravity wave pair $-\zeta\pm\mathrm i\tilde\omega$ and the Marangoni pair $s_\pm$ of \eqref{eq:marangoni-poles}. The contour consists of the Bromwich line (1), the large arcs (2, 8), the banks of the cut (3, 4 above; 6, 7 below) and the small circle (5) around $s_{b,l}$. Along $x_l<x<x_u$ only $m\LO$ is discontinuous; along $x>x_u$ both $m\LO$ and $m\UP$ are.}
\label{fig:bromwich-contour}
\end{figure}
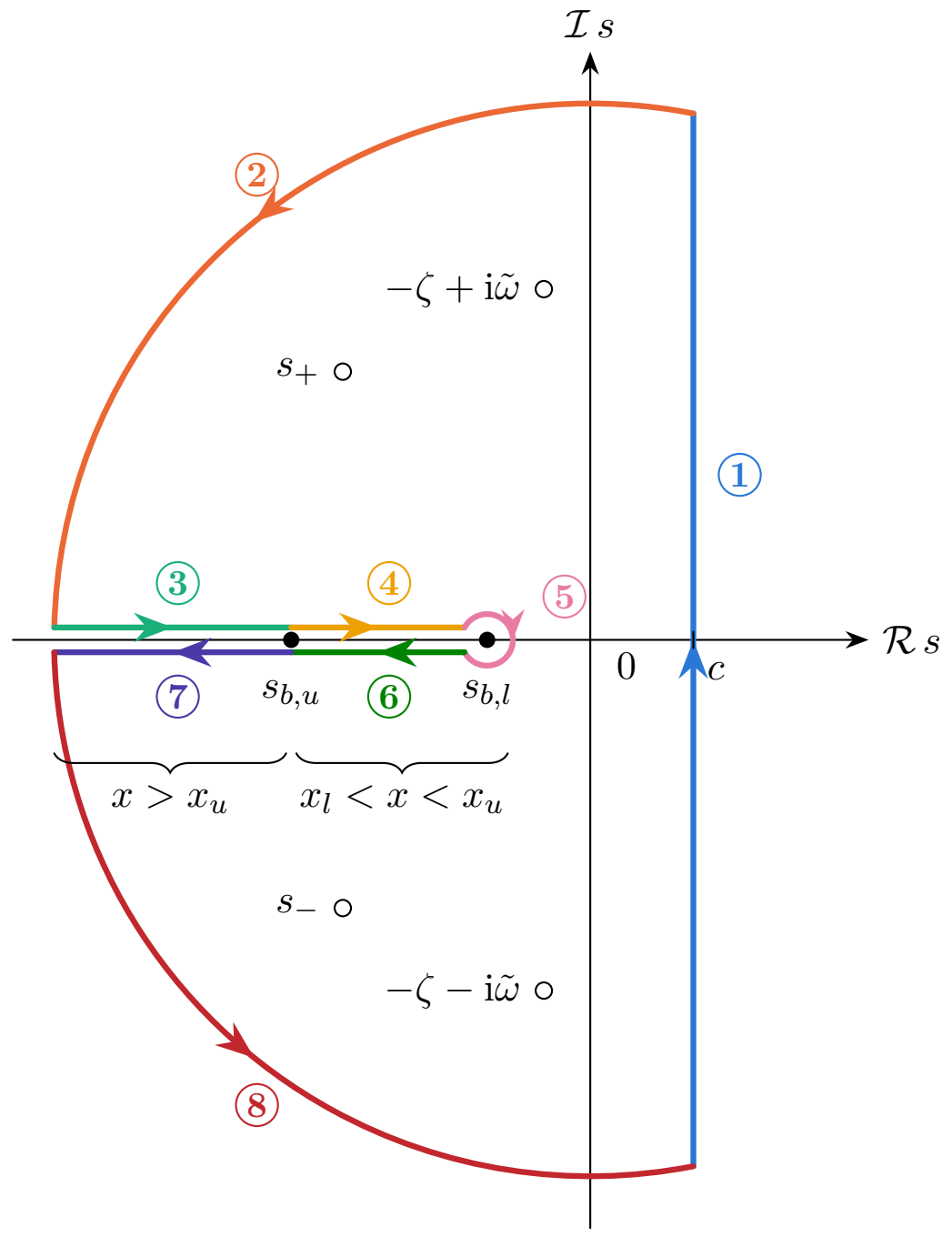

This appendix derives the inversion formula \eqref{eq:inversion} for the transformed concentration $\Gb(k,s)$; the argument is identical for $\eb(k,s)$. The wavenumber $k$ is fixed throughout. Write
\begin{equation}
  x_l=-s_{b,l}=\frac{k^2}{\Rey},\qquad x_u=-s_{b,u}=\frac{k^2\mu_r}{\Rey\,\rho_r},
  \label{eq:app-xlu}
\end{equation}
which are the inverse viscous diffusion times of the mode in the lower and upper fluid, in units of $t^*$. For definiteness we take $x_l<x_u$, i.e.\ $\mu_r/\rho_r>1$, the upper fluid having the larger kinematic viscosity, as for the gas-liquid pair and the liquid pairs $\phi=1$ and $2$ of this paper; for the pair $\phi=0.5$, $\mu_r/\rho_r=0.31$, and the roles of $m\LO$ and $m\UP$, and of $x_l$ and $x_u$, are exchanged, the cut then starting at $x_{\min}=x_u$. For a single fluid $m\UP$ is absent and the cut starts at $s_{b,l}$.

\emph{Branches.} The square roots \eqref{eq:m} are written as
\begin{equation}
  m\LO(s)=\Rey^{1/2}(s+x_l)^{1/2},\qquad m\UP(s)=\Big(\frac{\Rey\,\rho_r}{\mu_r}\Big)^{1/2}(s+x_u)^{1/2},\qquad -\pi<\arg(s+x_j)\le\pi,
  \label{eq:app-physical-branch}
\end{equation}
so that $\mathcal R\,m_j>0$ everywhere off the cuts, which run from $s_{b,j}$ to $-\infty$ along the negative real axis. This is the physical sheet, on which the vortical parts of \eqref{eq:bulk} decay away from the interface. On the two banks of the cut, $s=-x\pm\mathrm i0$ with $x>0$,
\begin{equation}
  m\LO^\pm(x)=
  \begin{cases}\big(k^2-\Rey\,x\big)^{1/2}, & x<x_l,\\[1mm]
  \pm\mathrm i\big(\Rey\,x-k^2\big)^{1/2}, & x>x_l,\end{cases}
  \qquad
  m\UP^\pm(x)=
  \begin{cases}\big(k^2-\Rey\,\rho_r x/\mu_r\big)^{1/2}, & x<x_u,\\[1mm]
  \pm\mathrm i\big(\Rey\,\rho_r x/\mu_r-k^2\big)^{1/2}, & x>x_u.\end{cases}
  \label{eq:mbranches}
\end{equation}
The transforms are therefore continuous across the negative real axis for $0<x<x_l$, only $m\LO$ jumps for $x_l<x<x_u$, and both $m\LO$ and $m\UP$ jump for $x>x_u$.

\emph{Behaviour at the origin and at infinity.} At $s=0$ the quantities $\Aa$, $\B$ and $\QQ$ of \eqref{eq:N-defs} vanish linearly in $s$, $\Nc$ vanishes linearly and $\NN$ tends to a non-zero limit, so that the four transfer functions of \eqref{eq:Gamma-sol}-\eqref{eq:eta-sol} are finite at the origin and $s=0$ is not a singularity. As $|s|\to\infty$ on the physical sheet, $\Gb\simeq\Gh_0/s$ and $\eb\simeq\eh_0/s$, while the cross terms decay faster; we have checked both limits numerically for the parameter sets of \S\ref{sec:single}.

\emph{Contour.} Let $\Phi(s)=\Gb(k,s)\,\mathrm e^{s\tr}$. The Bromwich integral
\begin{equation}
  \Gh(k,\tr)=\frac{1}{2\pi\mathrm i}\int_{c-\mathrm i\infty}^{c+\mathrm i\infty}\Phi(s)\,\mathrm ds,\qquad c>0,
  \label{eq:app-bromwich}
\end{equation}
is closed in the left half-plane by the contour of figure~\ref{fig:bromwich-contour}, which consists of the Bromwich line (1), the arcs (2) and (8) of a circle of radius $R$, the upper banks (3) for $x>x_u$ and (4) for $x_l<x<x_u$, the small circle (5) of radius $\delta$ about $s_{b,l}$, and the lower banks (6) and (7). No detour is needed at $s_{b,u}$, because the cut of $m\LO$ continues to the right of it and the contour simply runs along the bank from (3) into (4), and from (6) into (7). Denoting by $\mathcal J_i$ the contribution of piece $i$ divided by $2\pi\mathrm i$,
\begin{equation}
  \mathcal J_1+\mathcal J_2+\dots+\mathcal J_8=\sum_n\mathrm{Res}_{s_n}\big[\Gb\big]\,\mathrm e^{s_n\tr},
  \label{eq:app-residue-theorem}
\end{equation}
where $s_n$ are the zeros of $\NN$ on the physical sheet. They all have $\mathcal R\,s_n<0$, since the system is stable, so that any $c>0$ may be chosen, and they are real or occur in complex-conjugate pairs; for a simple pole the residue is $\Nc(s_n)\Gh_0/[s_n\NN'(s_n)]$. At the critical conditions at which two real poles merge into a complex pair, the pole is double and its residue involves the derivative of $\Nc\mathrm e^{s\tr}/s$, giving a term proportional to $\tr\,\mathrm e^{s_n\tr}$; the time-domain solutions of this paper are obtained by numerical inversion of the Laplace transform, which does not require the poles and is unaffected by their multiplicity.

\emph{Arcs and small circle.} Since $\Gb=O(s^{-1})$ on the arcs, Jordan's lemma gives $\mathcal J_2,\mathcal J_8\to0$ as $R\to\infty$ for $\tr>0$. On the small circle $s=s_{b,l}+\delta\mathrm e^{\mathrm i\vartheta}$, $\vartheta$ running from $\pi$ to $-\pi$, $m\LO=O(\delta^{1/2})$ while $\Gb$ remains bounded, so that $\mathcal J_5=O(\delta)\to0$.

\emph{Banks of the cut.} On the upper bank $s=x\,\mathrm e^{\mathrm i\pi}$ and on the lower bank $s=x\,\mathrm e^{-\mathrm i\pi}$, with $\mathrm ds=-\mathrm dx$ on both, so that
\begin{equation}
\begin{aligned}
  \mathcal J_3+\mathcal J_4&=\frac{1}{2\pi\mathrm i}\int_{x_l}^{\infty}\Gb^+(k,-x)\,\mathrm e^{-x\tr}\,\mathrm dx,\\
  \mathcal J_6+\mathcal J_7&=-\frac{1}{2\pi\mathrm i}\int_{x_l}^{\infty}\Gb^-(k,-x)\,\mathrm e^{-x\tr}\,\mathrm dx,
\end{aligned}
\label{eq:app-I-banks}
\end{equation}
where $\Gb^\pm(k,-x)$ is $\Gb$ evaluated with $m\LO^\pm$ and $m\UP^\pm$ of \eqref{eq:mbranches}. The coefficients of $\NN$, $\Nc$ and $\QQ$ are real, and $m_j^-=(m_j^+)^*$, so that $\Gb^-=(\Gb^+)^*$ and the four bank integrals combine into
\begin{equation}
  \mathcal J_3+\mathcal J_4+\mathcal J_6+\mathcal J_7=\frac1\pi\int_{x_l}^\infty\mathrm e^{-x\tr}\,\mathcal I\big[\Gb^+(k,-x)\big]\,\mathrm dx .
  \label{eq:app-banks-sum}
\end{equation}

\emph{Result.} With $\mathcal J_2,\mathcal J_5,\mathcal J_8\to0$, \eqref{eq:app-residue-theorem} and \eqref{eq:app-banks-sum} give $\Gh(k,\tr)=\mathcal J_1$ as
\begin{equation}
  \Gh(k,\tr)=\sum_n\mathrm{Res}_{s_n}\big[\Gb\big]\,\mathrm e^{s_n\tr}-\frac1\pi\int_{x_l}^{\infty}\mathrm e^{-x\tr}\,\mathcal I\big[\Gb^+(k,-x)\big]\,\mathrm dx ,
  \label{eq:app-final-inversion}
\end{equation}
which is \eqref{eq:inversion} with $x_{\min}=x_l$. In practice the cut integral is evaluated in the two pieces $x_l<x<x_u$, where $m\LO^+$ is imaginary and $m\UP^+$ real, and $x>x_u$, where both are imaginary. At late times the cut integral is dominated by the neighbourhood of $x=x_l$, where $\mathcal I\,\Gb^+$ vanishes as $(x-x_l)^{1/2}$, and decays as $\mathrm e^{-x_l\tr}\tr^{-3/2}$; in the large-$\Rey$ limit, where $x_l\to0$, this becomes the algebraic $\tr^{-3/2}$ decay of \citet{Shardt2016oscillatory} that appears in \eqref{eq:theta-asymptotes}.

\section{Expansions of the dispersion function}
\label{sec:appB}

\emph{Stokes limit.} With $m\LO=k+\Rey s/2k-\Rey^2s^2/8k^3+\ldots$ and $m\UP=k+\delta^2\Rey s/2k-\delta^4\Rey^2s^2/8k^3+\ldots$, where $\delta^2=\rho_r/\mu_r$, the quantities \eqref{eq:N-defs} expand as
\begin{equation}
\begin{gathered}
  \NN=-\frac{\Rey\,\rho_r}{4\mu_r}\big[G+2k(1+\mu_r)s\big]\big[k+2(1+\mu_r)s\big]+O(\Rey^2),\\
  \Nc=-\frac{\Rey\,\rho_r\,s}{2\mu_r}(1+\mu_r)\big[G+2k(1+\mu_r)s\big]+O(\Rey^2),\qquad
  \QQ=-\frac{\Rey^2\rho_r\,s^3(1-\rho_r)}{8k^3\mu_r}+O(\Rey^3).
\end{gathered}
\label{eq:app-stokes}
\end{equation}
The $O(\Rey)$ part of $\QQ$ vanishes identically: the viscous normal stress of the boundary layers, $2k(1-\mu_r)(k-m\LO)(k-m\UP)/\Rey\simeq(1-\mu_r)\delta^2\Rey s^2/2k$, cancels the pressure term $s\B\simeq-(\delta^2-\rho_r)\Rey s^2/2k$ because $\mu_r\delta^2=\rho_r$. Substituting \eqref{eq:app-stokes} in \eqref{eq:Gamma-sol}-\eqref{eq:eta-sol} gives \eqref{eq:stokes-sol} and \eqref{eq:stokes-ORe}; for the single fluid the common factor $k-m\UP$ is cancelled first.

\emph{Large Reynolds number.} Write $L=(\Rey s)^{1/2}$ and $\delta=(\rho_r/\mu_r)^{1/2}$, so that $m\LO=(L^2+k^2)^{1/2}$, $m\UP=(\delta^2L^2+k^2)^{1/2}$ and $\rho_r+\delta=\delta(1+\phi)$. The square roots must be expanded consistently, $m\LO=L+k^2/2L+\ldots$ and $m\UP=\delta L+k^2/2\delta L+\ldots$, because their corrections, multiplied by the $L^3$ and $L^2$ terms, contribute at $O(L)$ and $O(1)$. The result is
\begin{equation}
\begin{aligned}
  \NN={}&-\delta(1+\phi)(1+\rho_r)\,sL^3+\big[k(1-\rho_r)^2s-k^2\delta(1+\rho_r)\big]L^2\\
  &-kG\delta(1+\phi)\,L+k^2G(1+\rho_r)-\frac{k^3G\delta}{s}+N_1L+N_0,
\end{aligned}
\label{eq:app-largeRe}
\end{equation}
where $N_1=O(k^3,k^2s)$ and $N_0=O(k^4,k^3s,k^2s^2)$ collect the remaining terms of first and zeroth order in $L$; they are lengthy and, like the corresponding terms of $\Nc$, are given in the supplementary material. $\Nc$ has the same $L^3$ term, the $L^2$ term $k(1-\rho_r)^2sL^2$, and the $G$ terms $-kG\delta(1+\phi)L+k^2G(1+\rho_r)$; and $\QQ\simeq\delta(\phi-1)sL$. At the Marangoni scale $s=O(k^{4/3}\Rey^{-1/3})$ the terms $sL^3$ and $k^2\delta L^2$ are of the same order, $k^{10/3}\Rey^{2/3}$, the terms in $G$ are of the same order when $G=O(k^{5/3}\Rey^{1/3})$, which is the distinguished limit \eqref{eq:crossing}, and the remaining terms are smaller by powers of $k/L$. Retaining the four dominant terms and using $L^2=\Rey s$ and $kG=(1+\rho_r)\Rey\tilde\omega_0^2$ gives \eqref{eq:largeRe-N}; the leading term of $\QQ$ gives \eqref{eq:eta-largeRe}, and the expansion of \eqref{eq:theta-largeRe} and \eqref{eq:eta-largeRe} for large $s$ gives the early-time laws of \S\ref{sec:largeRe}.

The damping \eqref{eq:damping} and the residue \eqref{eq:residue} require the next orders. The damping follows from the terms of relative order $k/L$, evaluated at $s=\mathrm i\tilde\omega_0$ and divided by the derivative of the leading factor; this is the weak-viscosity expansion of the dispersion relation of \citet{Rajan2020resonance}. The residue at the wave pole vanishes at leading order, because the wave factor cancels between $\Nc$ and $\NN$, and requires all terms of \eqref{eq:app-largeRe} through $O(1)$; with only the terms through $O(L)$ the residue is in error by a factor close to four. Both results were checked against the exact poles and residues. For a gas-liquid interface at $q=1$ the ratio of the exact damping to \eqref{eq:damping} is $1.15$, $1.07$, $1.03$ and $1.01$ at $\Rey=10^4$, $10^5$, $10^6$ and $10^7$, and at $q=2$ it is $0.99$, $1.005$, $1.005$ and $1.003$; the difference decreases as $\Rey^{-1/3}$, as $k/L$ does at fixed $q$, since then $E\propto\Rey^{-1/3}$, $s\sim\tilde\omega_0\propto\Rey^{-1/3}$ and $L=(\Rey s)^{1/2}\propto\Rey^{1/3}$. The residue computed from the consistent expansion \eqref{eq:app-largeRe} through $O(1)$ agrees with \eqref{eq:residue} to within $1\,\%$ for $0.4\le q\le2.5$ at $E=10^{-3}$, and the exact residue agrees with it to within $0.1\,\%$ (figure~\ref{fig:residue}). Wherever numbers are quoted in the text, the exact poles and residues are used.

\textbf{Acknowledgements:}	Institute Fellowship to Mr. Abhinandan Mohanty by Indian Institute of Technology, Roorkee is deeply acknowledged. PKF acknowledges funding support provided by Indian Institute of Technology, Roorkee (grant No. FIG101039). We gratefully acknowledge the Anusandhan National Research Foundation (ANRF grant ANRF/ARG/2025/003821/ENS) for supporting fundamental investigation of physicochemical phenomena on interfacial flows.

\textbf{Supplementary material.} Supplementary material, giving the details of the numerical method, its verification and the coefficients of the large-Reynolds-number expansion, is available as an ancillary document.

\textbf{Declaration of interests.} The authors report no conflict of interest.

\textbf{Data availability statement.} The scripts that evaluate the linear solution, the simulation drivers and the data underlying the figures are available from the corresponding author on reasonable request and will be deposited in a public repository on publication.

\bibliographystyle{plainnat}
\bibliography{references_waves}

\end{document}